\documentclass[useAMS,usenatbib,usegraphicx]{mn2e}

\def\aj{AJ}%
\def\apj{ApJ}%
\def\apjl{ApJ}%
\def\apjs{ApJS}%
\def\aap{A\&A}%
\def\aaps{A\&AS}%
\def\mnras{MNRAS}%
\newcommand{\sla}{\mathrm{SL}_{0.5}}
\newcommand{\slb}{\mathrm{SL}_{0.75}}
\newcommand{\kpc}{\mathrm{kpc}}
\newcommand{\kms}{\mathrm{km/s}}
\newcommand{\reff}{r_\mathrm{eff}}
\newcommand{\rein}{r_\mathrm{Ein}}
\newcommand{\ieff}{\langle I \rangle_\mathrm{eff}}
\newcommand{\siggal}{\sigma_\mathrm{eff}}
\newcommand{\sigeff}{\langle \Sigma \rangle_\mathrm{eff}}
\newcommand{\sigdyn}{\langle \Sigma \rangle_{\ast,\mathrm{dyn}}}
\newcommand{\sigtot}{\langle \Sigma \rangle_{\mathrm{tot,dyn}}}
\newcommand{\iun}{\mathrm{L}_\odot \mathrm{pc}^{-2}}
\newcommand{\mun}{\mathrm{M}_\odot \mathrm{pc}^{-2}}
\newcommand{\gun}{(\kms)^2 \, \kpc \, \mathrm{M}_\odot^{-1}}
\newcommand{\mlun}{\mathrm{M}_\odot/\mathrm{L}_{R,\odot}}
\newcommand{\msun}{\mathrm{M}_\odot}

\newcommand{\rh}{r_h}
\newcommand{\Gyr}{\mathrm{Gyr}}
\newcommand{\mlssp}{\Upsilon_\mathrm{ssp}}
\newcommand{\mlkroupa}{\Upsilon_\mathrm{Krou}}
\newcommand{\mlsalp}{\Upsilon_\mathrm{Salp}}
\newcommand{\ratml}{\mldyn/\mlkroupa}
\newcommand{\mtot}{M_\mathrm{tot,dyn}}
\newcommand{\mtotgal}{M_\mathrm{tot,gal}}
\newcommand{\mdm}{M_\mathrm{DM,dyn}}
\newcommand{\mein}{M_\mathrm{Ein}}

\newcommand{\mdyn}{M_{\ast,\mathrm{dyn}}}
\newcommand{\msc}{M_{\ast,\mathrm{sc}}}

\newcommand{\dmfrac}{f_\mathrm{DM,dyn}}
\newcommand{\dmfracgal}{f_\mathrm{DM,gal}}
\newcommand{\dmfrack}{f_\mathrm{DM,Krou}}
\newcommand{\mldyn}{\Upsilon_{\ast,\mathrm{dyn}}}
\newcommand{\mlsc}{\Upsilon_{\ast,\mathrm{sc}}}
\newcommand{\mlgal}{\Upsilon_{\ast,\mathrm{gal}}}
\newcommand{\upstot}{\Upsilon_{\mathrm{tot,dyn}}}
\newcommand{\mlcor}{f_\ast \Upsilon_\mathrm{dyn}}
\newcommand{\mltot}{\mtot/L}
\newcommand{\avml}{\langle \Upsilon \rangle}

\newcommand{\rhodm}{\rho_\mathrm{DM}}
\newcommand{\rhodmk}{\rho_\mathrm{DM,Krou}}
\newcommand{\deltadm}{\Delta f_\mathrm{DM}}
\newcommand{\deltaups}{\Delta \Upsilon}

\title[Stellar IMF and luminous/dark matter in early-type
galaxies]{Dynamical masses of early-type galaxies: a comparison to
  lensing results and implications for the stellar IMF and the
  distribution of dark matter} \author[J. Thomas et
al.]{J. Thomas$^{1,2}$\thanks{E-mail: jthomas@mpe.mpg.de},
  R. P. Saglia$^{1,2}$, R. Bender$^{1,2}$, D. Thomas$^{3}$,
  K. Gebhardt$^{4}$, \newauthor J. Magorrian$^{5}$,
  E.~M. Corsini$^{6}$, G. Wegner$^{7}$
  and S. Seitz$^{1,2}$\\
  $^{1}$Universit\"atssternwarte M\"unchen, Scheinerstra\ss e 1, D-81679 M\"unchen, Germany\\
  $^{2}$Max-Planck-Institut f\"ur Extraterrestrische Physik, Giessenbachstra\ss e, D-85748 Garching, Germany\\
  $^{3}$Institute of Cosmology and Gravitation, University of Portsmouth, Dennis Sciama Building, Burnaby Road, Portsmouth, PO1 3FX, UK\\
  $^{4}$Department of Astronomy, University of Texas at Austin, C1400, Austin, TX78712, USA\\
  $^{5}$Theoretical Physics, Department of Physics, University of Oxford, 1 Keble Road, Oxford U.K., OX1 3NP\\
  $^{6}$Dipartimento di Astronomia, Universit\`a di Padova, vicolo dell'Osservatorio 2, I-35122 Padova, Italy\\
  $^{7}$Department of Physics and Astronomy, 6127 Wilder Laboratory,
  Dartmouth College, Hanover, NH 03755-3528, USA}
\begin{document}

\date{Accepted 1988 December 15. Received 1988 December 14; in original form 1988 October 11}

\pagerange{\pageref{firstpage}--\pageref{lastpage}} \pubyear{2002}

\maketitle

\label{firstpage}

\begin{abstract}
  This work aims to study the distribution of luminous and dark matter
  in Coma early-type galaxies.  Dynamical masses obtained under the
  assumption that mass follows light do not match with the masses of
  strong gravitational lens systems of similar velocity dispersions.
  Instead, dynamical fits with dark matter halos are in good agreement
  with lensing results.  We derive mass-to-light ratios of the stellar
  populations from Lick absorption line indices, reproducing well the
  observed galaxy colours.  Even in dynamical models with dark matter
  halos the amount of mass that follows the light increases more
  rapidly with galaxy velocity dispersion than expected for a constant
  stellar initial mass function (IMF). While galaxies around $\siggal
  \approx 200 \, \kms$ are consistent with a Kroupa IMF, the same IMF
  underpredicts luminous dynamical masses of galaxies with $\siggal
  \approx 300 \, \kms$ by a factor of two and more. A systematic
  variation of the stellar IMF with galaxy velocity dispersion could
  explain this trend with a Salpeter IMF for the most massive
  galaxies. If the IMF is instead constant, then some of the dark
  matter in high velocity dispersion galaxies must follow a spatial
  distribution very similar to that of the light. A combination of
  both, a varying IMF and a component of dark matter that follows the
  light is possible as well. For a subsample of galaxies with old
  stellar populations we show that the tilt in the fundamental plane
  can be explained by systematic variations of the total (stellar +
  dark) mass inside the effective radius. We tested commonly used mass
  estimator formulae, finding them accurate at the $20-30 \, \%$
  level.
\end{abstract}

\begin{keywords}
galaxies: elliptical and lenticular, cD -- 
galaxies: kinematics and dynamics --- galaxies: structure
\end{keywords}

\section{Introduction}
\label{sec:intro}
The masses of galaxies are revealed by the gravitational interaction
of their matter constituents, e.g. by stellar or gas kinematics or
gravitational lensing effects. The caveat is that these effects
rigorously constrain only the total amount of gravitating mass. The
decomposition into luminous and dark matter relies on further
assumptions. For example, in dynamical studies of early-type galaxies
it is commonly assumed that the stellar mass distribution follows the
light. Any radial increase in the mass-to-light ratio -- if needed to
explain the observed velocities of the stars -- is attributed to dark
matter.

The limitation of this approach becomes clear if one imagines a galaxy
dark matter halo that follows the light distribution
exactly. Discriminating between luminous and dark matter according to
their potentially different radial distributions would fail in this
case. In fact, there would be no direct way to dynamically unravel the
relative contributions of luminous and dark matter to the galaxy
mass. It is unlikely that a real galaxy halo follows the light
distribution exactly, yet it exemplifies the intrinsic degeneracies in
any mass decomposition. On top of this, even the most advanced
present-day dynamical modelling techniques rely on symmetry
assumptions and, often, also on the assumption of a steady-state
dynamical system. If the symmetry assumptions are strongly violated,
dynamical mass-to-light ratios can be biased by a factor of up to two
\citep{Tho07A}.

To crosscheck the validity of the assumptions in dynamical modelling
it is important to compare the resulting masses with other independent
methods. {\it Total} masses can be most directly compared to results
from gravitational lensing, which is the first goal of this paper.

An examination of the mass {\it decomposition} requires the
investigation of a galaxy's stellar population. Since the latter
provides an independent measure of the stellar mass-to-light ratio,
the comparison of dynamical and stellar-population masses yields
further constraints on the dark matter content. In the above mentioned
case, for example, the dynamical mass-to-light ratio $\mldyn$ would
exceed the corresponding stellar population value, indicating that
some fraction of the galaxy's mass is actually dark matter.

Unfortunately, stellar population models do not provide unique stellar
mass-to-light ratios. They suffer from an incomplete knowledge of the
initial stellar mass function (IMF), as well as age-metallicity
degeneracies. Observations in and around our own Galaxy indicate that
the IMF slope flattens below $0.5 \, \msun$ \citep{Sca86,Kro01}.
Recent spectroscopic observations of massive ellipticals in the
near-infrared point towards the low-mass slope of the IMF in these
galaxies being steeper \citep{Dok10}. Yet, until now the IMF in
distant galaxies with unresolved stellar populations remains largely
uncertain. This translates into a significant indeterminacy of
population mass-to-light ratios: the steeper the slope at the low-mass
end, the higher the population mass-to-light ratio. Hence, the stellar
population approach is not directly conclusive as a probe of the mass
decomposition in dynamical models.

%%%%%%%%%%%%%%%%%%%%%%%%%%%%%%%%%%%%%%%%%%
% SSP Data
%%%%%%%%%%%%%%%%%%%%%%%%%%%%%%%%%%%%%%%%%%
\begin{figure*}\centering
\begin{minipage}{166mm}
\includegraphics[width=144mm,angle=0]{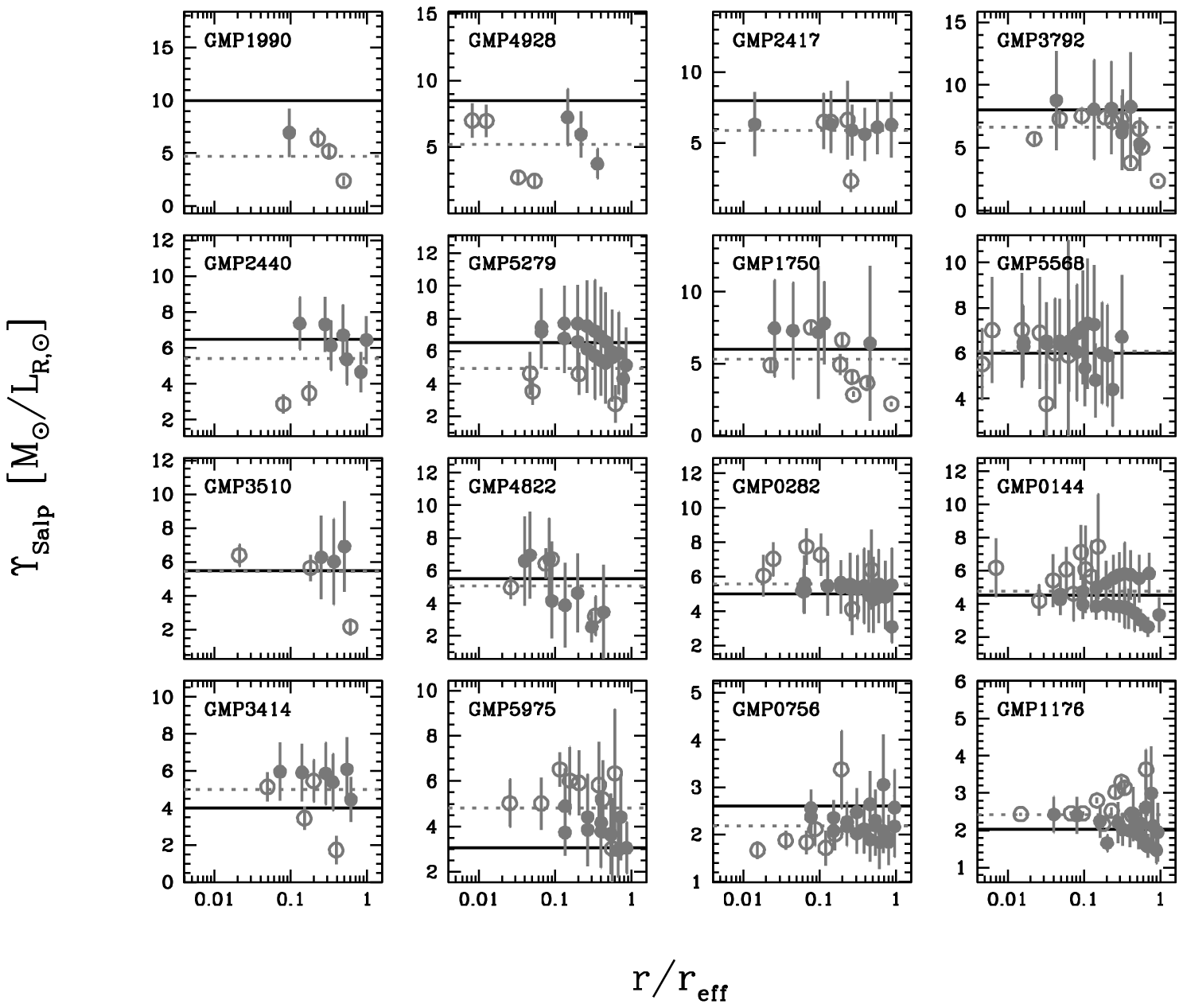}
\caption{Stellar population mass-to-light ratios $\mlsalp$ (Salpeter
  IMF) as function of radius along the major-axis (filled circles) and
  minor-axis (open circles). Note that for some galaxies there
    are systematic differences between the two sides of a slit which
    are slightly larger than the statistical errors (e.g. along the
    major-axis of GMP0144). The dotted line in each panel is the
  light-weighted average of $\mlsalp$ within $r<\reff$, the solid line
  corresponds to the stellar mass-to-light ratio $\mldyn$ from
  dynamical models. From top-left to bottom-right galaxies are plotted
  in order of decreasing $\mldyn$.}
\label{fig:salpeter_data}
\end{minipage}
\end{figure*}

Conversely, if dynamical stellar mass determinations were free of
ambiguities with respect to a dark matter contamination then they
could serve as a measure for the slope of the IMF in distant
galaxies. The method would be to tweak the IMF in the stellar
population models until agreement with the dynamical stellar masses is
achieved.

The ambiguities in both, dynamical stellar masses as well as stellar
population models, make neither of the approaches directly
applicable. Nevertheless, comparing dynamical with stellar population
models is important to (1) narrow down the dynamically plausible range
of possible IMFs and to (2) delimit the range of dark matter fractions
compatible with observed stellar population properties. This is the
second goal of the present paper in which we compare dynamical and
stellar population masses in a sample of 16 Coma early-type galaxies.

Our dynamical models account for both the full variety of possible
orbit configurations in axisymmetric, flattened galaxies as well as
for dark matter. In this respect we extend previous
studies. \citet{Cap06}, for example, used a similar modelling
technique to compare dynamical and stellar population masses in the
SAURON sample, but they did not consider dark matter explicitly in
their dynamical models.  The justification for this neglect was the
expected insignificance of dark matter in the central galaxy regions
probed by the SAURON observations ($r_\mathrm{obs,max} \la
\reff$). However, measuring only the sum of luminous and dark mass
makes the comparison with stellar population models potentially
uncertain.  \citet{Nap10} analysed a large sample of early-type
galaxies taking into account the contribution from dark matter, but
their models do not account for galaxy flattening and rotation.

A subsample of the Coma galaxies was recently analysed by
\citet{Gri10}. While they used multi-band photometry to derive stellar
population properties, our approach is to measure ages, metallicities
and [$\alpha$/Fe] ratios from Lick indices to reduce potential biases
in population parameters.  Likewise, the analysis of SLACS galaxies by
\citet{Tre10}, combining constraints from gravitational lensing and
stellar dynamics, relied on multi-band photometry for the stellar
population part.

Mass-to-light ratios of early-type galaxies are of particular interest
to understand the tilt of the fundamental plane. Virial relations
imply that the effective surface brightness $\ieff$, the effective
radius $\reff$ and the central velocity dispersion $\sigma_0$ in hot
stellar systems are not independent of each other.  This is revealed
by the fundamental plane of early-type galaxies \citep{Djo87,Dre87}.
Yet, the observed fundamental plane is tilted with respect to the
simple case of a virialised, homologous family of dynamical
systems. This tilt can reflect (1) systematic variations of the
luminosity distribution (e.g. \citealt{Sag93,Tru04}), (2) systematic
variations of the orbital structure (e.g. \citealt{Cio96}) or (3)
systematic variations of the mass-to-light ratio, as a result of
varying stellar populations and/or dark matter distributions
(e.g. \citealt{Ren93}).  Understanding these variations allows a
deeper insight into the formation process of early-type galaxies
\citep{Ben92}.

Most of the above mentioned effects can be factored out if additional
information about the stellar populations and/or the mass
distributions are available. Aperture spectroscopy is one way to
measure stellar population properties. By assuming simple scaling laws
it can also provide estimates for dynamical masses, such that the
contributions of stellar population and dynamical effects on the
fundamental plane tilt can be disentangled (e.g.
\citealt{Hyd09,Gra10}). More reliable constraints come from radially
resolved spectroscopy and detailed dynamical (or lensing) models of
galaxies (e.g. \citealt{Cap06,Bol07}). The third goal of this paper is
to follow the latter approach and to use the specific information
contained in our two-component dynamical models for further
investigations upon the origin of the fundamental plane tilt.

The paper is organised as follows. Sec.~\ref{data:review} reviews the
galaxy sample and models. in Sec.~\ref{sec:projmass} we compare
projected masses from dynamical models and from gravitational lensing
against each other. Sec.~\ref{sec:ml} deals with the comparison of
luminous dynamical and stellar population masses. The dark matter
distribution is analysed in Sec.~\ref{sec:dm} and implications for the
tilt of the fundamental plane are addressed in Sec.~\ref{sec:fp}. The
paper is summarised in Sec.~\ref{sec:sum}.

\section{Galaxy sample and modelling}
\label{data:review}

The sample analysed in this paper comprises 16 Coma early-type
galaxies in the luminosity range between $M_B = -19.88$ and $M_B = -
22.26$ (eight giant ellipticals, eight lenticular/intermediate type
galaxies). It is almost identical to the sample presented in
\citet{Tho09}. Only the galaxy GMP3958 has been omitted for its strong
gas emission, which hampers reliable stellar population modelling.

For the dynamical analysis of each galaxy a composite of ground-based
and HST photometry has been used. Stellar absorption line data for the
kinematics and Lick indices come from various long-slit spectra, at
least along the major and minor axes, but in many cases covering other
position angles as well. The spectra extend to $1-4\,\reff$. Details
about the photometric and spectroscopic data have been published in
\citet{Meh00}, \citet{Weg02}, \citet{Cor08}, and \citet{Tho09}.

\subsection{Dynamical modelling}
\label{subsec:dynmod}
To the kinematic and photometric data we applied our implementation of
Schwarzschild's orbit superposition technique \citep{S79} for
axisymmetric potentials \citep{Ric88,Geb03,Tho04,Tho05}.  A detailed
description of the models is given in \citet{Tho07B}.  Most important
for this paper are the assumptions about the mass structure. We will
consider two sets of models. For the first set it is assumed that all
the mass follows the light, i.e.
\begin{equation}
\label{eq:mfl}
\rho = \Upsilon \times \nu
\end{equation}
where $\nu$ is the three-dimensional luminosity density. By
construction, the mass-to-light ratio $\Upsilon$ here includes the
contribution of both the stellar and dark mass of a galaxy. It is not
known in advance and obtained by a $\chi^2$-minimisation with respect
to the kinematical observations. The best-fit $\Upsilon$ of
one-component (i.e. self consistent) models will be referred to as
$\mlsc$ in the remainder of this paper and can be found in
Tab.~\ref{tab:dmfrac}.

For the second, and standard set of models we assume a mass density of
the form
\begin{equation}
\label{eq:massdecomp}
\rho = \Upsilon \times \nu + \rho_\mathrm{DM}.
\end{equation}
The first component again follows the light, while the second one,
$\rhodm$, accounts for dark matter. Eq.~\ref{eq:massdecomp} is
designed to separate the contributions of luminous and dark matter to
the total mass of a galaxy. Our basic assumption is that the best-fit
$\Upsilon$ of two-component models represents only the stellar mass of
a galaxy and we will refer to it as the dynamical stellar
mass-to-light ratio $\mldyn$ in the following.  Strictly speaking,
$\mldyn$ measures all the mass which follows the light distribution,
be it stellar or be it dark matter. In this respect $\mldyn$ provides
only an upper limit for the stellar mass. As soon as there are other
mass components which -- for whatever reason -- follow the light, they
will contribute to $\mldyn$ as well and the actual galaxy stellar
mass-to-light ratio will be smaller than $\mldyn$.  A more detailed
discussion upon this issue will be given in
Secs.~\ref{subsec:mldynerr} and \ref{subsec:lightdm}.

According to equation (\ref{eq:massdecomp}), the cumulative
(spherical) dark matter fraction $\dmfrac$ of the models inside radius
$r$ is
\begin{equation}
\label{eq:dmfrac}
\dmfrac(r) \equiv \frac{\mdm(r)}{\mtot(r)} = \frac{\mdm(r)}{\mldyn \, L(r) + \mdm(r)},
\end{equation}
where $\mtot$, $\mdm$ and $L$ stand for the cumulative total mass,
dark mass and light inside radius $r$. Tab.~\ref{tab:dmfrac} lists the
observationally derived dark matter fractions. Here we consider mostly
the region inside $\reff$, where the average dark matter fraction is
$\langle \dmfrac \rangle = 23 \pm 17 \, \%$, but our data reach out to
radii where the stellar and the dark matter density become equal
\citep{Tho07B}.

\begin{table*}
\begin{center}
\begin{minipage}{170mm}
\begin{tabular}{lc||cccccc}
\multicolumn{2}{c}{galaxy} &  $\mldyn$           &  $\mlsc$           & $\dmfrac$             & $\siggal$            & $\mlkroupa$       & $\mlsalp$\\      \\
GMP & NGC/IC               &  $[\mlun]$          &  $[\mlun]$          &                       & $[\kms]$             & $[\mlun]$         & $[\mlun]$\\      \\
(1) & (2)                  & (3)                 & (4)                 & (5)                   & (6)                  &   (7)             & (8)              \\
\hline                                                                                                                                                       
0144 & 4957                &  $4.50 \pm  0.50$   &  $7.00 $   & $0.30^{+0.10}_{-0.05} $ & $211.8 \pm 0.4$     & $3.03 \pm  0.17$  & $4.74 \pm  0.26$ \\ 
0282 & 4952                &  $5.00 \pm  0.50$   &  $6.50 $   & $0.25^{+0.13}_{-0.05} $ & $268.2 \pm 0.5$     & $3.57 \pm  0.22$  & $5.58 \pm  0.34$ \\ 
0756 & 4944                &  $2.60 \pm  0.20$   &  $3.00 $   & $0.11^{+0.05}_{-0.05} $ & $193.2 \pm 0.3$     & $1.40 \pm  0.06$  & $2.19 \pm  0.10$ \\ 
1176 & 4931                &  $2.00 \pm  0.20$   &  $2.50 $   & $0.31^{+0.05}_{-0.04} $ & $205.6 \pm 0.3$     & $1.54 \pm  0.04$  & $2.41 \pm  0.06$ \\ 
1750 & 4926                &  $6.00 \pm  0.75$   &  $7.00 $   & $0.15^{+0.47}_{-0.05} $ & $279.1 \pm 1.6$     & $3.41 \pm  0.50$  & $5.33 \pm  0.79$ \\
1990 & IC 843              &  $10.00 \pm  1.00$  &  $10.00 $  & $0.01^{+0.01}_{-0.01} $ & $281.3 \pm 1.1$     & $3.03 \pm  0.30$  & $4.74 \pm  0.47$ \\ 
2417 & 4908                &  $8.00 \pm  0.75$   &  $8.50 $   & $0.05^{+0.20}_{-0.04} $ & $211.0 \pm 1.7$     & $3.78 \pm  0.43$  & $5.91 \pm  0.67$ \\ 
2440 & IC 4045             &  $6.50 \pm  0.50$   &  $7.00 $   & $0.07^{+0.02}_{-0.02} $ & $225.9 \pm 0.9$     & $3.46 \pm  0.27$  & $5.41 \pm  0.42$ \\ 
3414 & 4871                &  $4.00 \pm  0.62$   &  $6.00 $   & $0.46^{+0.06}_{-0.27} $ & $169.6 \pm 1.3$     & $3.20 \pm  0.29$  & $4.99 \pm  0.45$ \\ 
3510 & 4869                &  $5.50 \pm  0.50$   &  $6.00 $   & $0.09^{+0.18}_{-0.05} $ & $177.7 \pm 1.7$     & $3.49 \pm  0.65$  & $5.45 \pm  1.02$ \\ 
3792 & 4860                &  $8.00 \pm  1.00$   &  $9.00 $   & $0.13^{+0.24}_{-0.05} $ & $284.0 \pm 1.9$     & $4.27 \pm  0.31$  & $6.66 \pm  0.48$ \\ 
4822 & 4841A               &  $5.50 \pm  1.00$   &  $6.50 $   & $0.51^{+0.14}_{-0.27} $ & $272.1 \pm 2.7$     & $3.24 \pm  0.33$  & $5.07 \pm  0.52$ \\
4928 & 4839                &  $8.50 \pm  2.00$   &  $10.00 $   & $0.32^{+0.35}_{-0.21} $ & $314.8 \pm 2.9$     & $3.34 \pm  0.46$  & $5.22 \pm  0.72$ \\ 
5279 & 4827                &  $6.50 \pm  0.50$   &  $7.00 $   & $0.10^{+0.21}_{-0.07} $ & $244.1 \pm 1.2$     & $3.16 \pm  0.45$  & $4.93 \pm  0.71$ \\ 
5568 & 4816                &  $6.00 \pm  1.00$   &  $7.00 $   & $0.53^{+0.15}_{-0.25} $ & $233.4 \pm 1.7$     & $3.91 \pm  0.33$  & $6.11 \pm  0.52$ \\ 
5975 & 4807                &  $3.00 \pm  0.50$   &  $4.00 $   & $0.29^{+0.05}_{-0.01} $ & $195.9 \pm 0.8$     & $3.07 \pm  0.24$  & $4.80 \pm  0.37$ \\ 
\hline 
\end{tabular}
\caption{Dynamical parameters of Coma galaxies.  (1, 2) galaxy
  identification (GMP from \citealt{GMP}). (3) best-fit dynamical
  $\mldyn$ ($R$-band) in models that explicitly account for dark
  matter (the quoted errors include all models that deviate by less
  than $\Delta \chi^2 \le 1$ from the best-fit model. For a more
  detailed discussion of the errors the reader is referred to
  \citealt{Tho05}.) (4) best-fit dynamical $\mlsc$ ($R$-band) assuming
  that all the mass follows the light.  (According to Fig.~2 in
  \citealt{Tho07B} the formal errors on $\mlsc$ are smaller than those
  of $\mldyn$. However, since the assumption that mass-follows-light
  does not yield satisfactory fits to the kinematics we don't give
  errors for $\mlsc$ here.)  (5) dark matter fraction $\dmfrac$ within
  $\reff$.  (6) galaxy velocity dispersion $\siggal$ (inside
  $\reff$). (7, 8) stellar population mass-to-light ratios for the
  Kroupa IMF ($\mlkroupa$; $R$-band) and Salpeter IMF ($\mlsalp$;
  $R$-band), respectively. The mass-to-light ratios are light-weighted
  averages within $\reff$. Only radii with a stellar-population age
  $\tau \le 14 \, \Gyr$ are considered.
\label{tab:dmfrac}}
\end{minipage}
\end{center}
\end{table*}

We probed two dark matter descriptions. Firstly, logarithmic halos
with a constant-density core of size $\rh$ and secondly NFW halos
\citep{Nav96}. The latter provide good fits to cosmological $N$-body
simulations and have a central logarithmic slope of $-1$.  Neither of
the two profiles includes baryonic halo contraction explicitly
(cf. the discussion in Sec.~\ref{subsec:lightdm}).  Coma galaxies are
in most cases better fit with logarithmic halos, but the significance
over NFW halo profiles is marginal. The steeper central slope of NFW
halos implies a higher central dark matter density. But even in cases
where NFW halos fit better, the innermost galaxy regions are still
dominated by the mass-component that follows the light (for details
see \citealt{Tho07B}). The assumptions about the halo-density profile
have therefore little influence on the best-fit stellar mass-to-light
ratio $\mldyn$ (typically, $\mldyn$ from logarithmic or NFW halos
differ by no more than $\Delta \mldyn \approx 0.5$; see also
Sec.~\ref{subsec:imfnorm}).

%%%%%%%%%%%%%%%%%%%%%%%%%%%%%%%%%%%%%%%%%%
% Ups_dyn/Ups_Krou vs sigma
%%%%%%%%%%%%%%%%%%%%%%%%%%%%%%%%%%%%%%%%%%
\begin{figure}\centering
\includegraphics[width=84mm,angle=0]{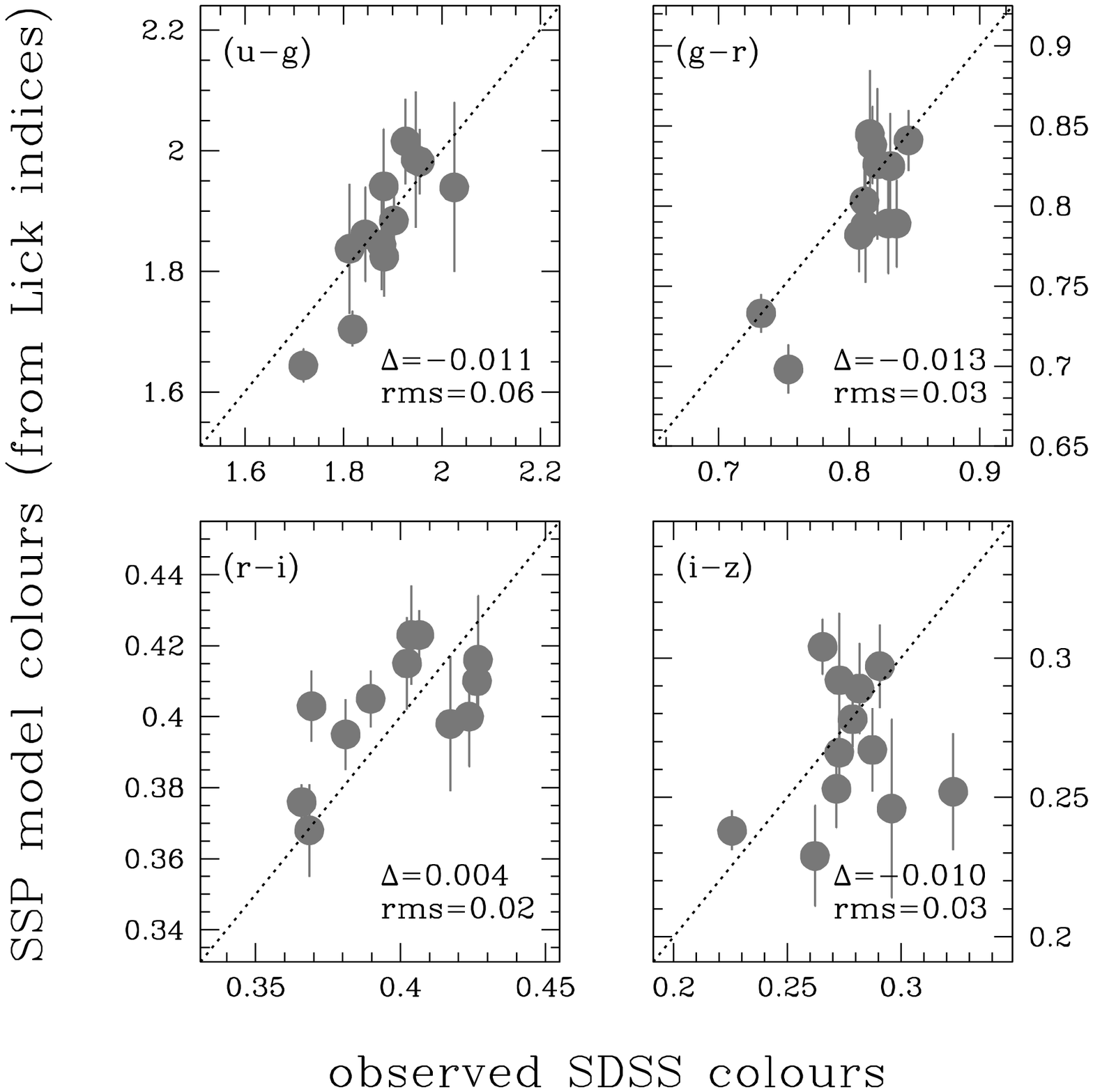}
\caption{Observed versus predicted colours for the subsample of Coma
  galaxies with SDSS photometry. Colours as indicated in each
  panel. Model predictions are plotted along the vertical axis, SDSS
  observations horizontally. All axes are in magnitudes.  The mean
  colour difference $\Delta$ and rms-scatter between models and
  observations (in magnitudes) is quoted in each panel.}
\label{fig:farben}
\end{figure}

%%%%%%%%%%%%%%%%%%%%%%%%%%%%%%%%%%%%%%%%%%
% Ups_dyn/Ups_Krou vs sigma
%%%%%%%%%%%%%%%%%%%%%%%%%%%%%%%%%%%%%%%%%%
\begin{figure}\centering
\includegraphics[width=84mm,angle=0]{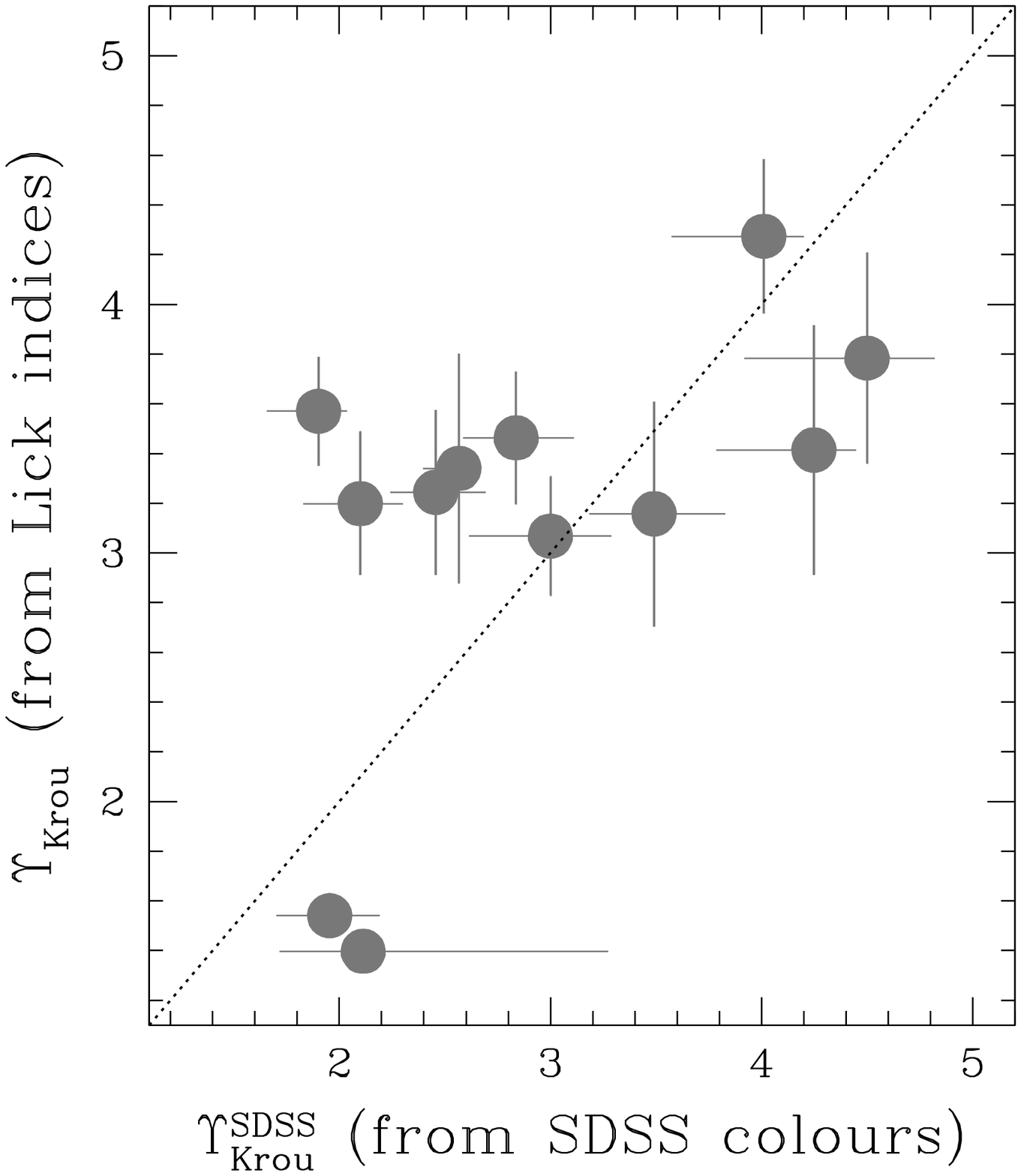}
\caption{Stellar population $\mlkroupa$ (Kroupa IMF) from SDSS colours
  (x-axis; cf. \citealt{Gri10}) and from Lick indices (y-axis;
  cf. Tab.~\ref{tab:dmfrac}).}
\label{fig:grillo}
\end{figure}

%%%%%%%%%%%%%%%%%%%%%%%%%%%%%%%%%%%%%%%%%%
% Ups_dyn/Ups_Krou vs sigma
%%%%%%%%%%%%%%%%%%%%%%%%%%%%%%%%%%%%%%%%%%
\begin{figure}\centering
\includegraphics[width=84mm,angle=0]{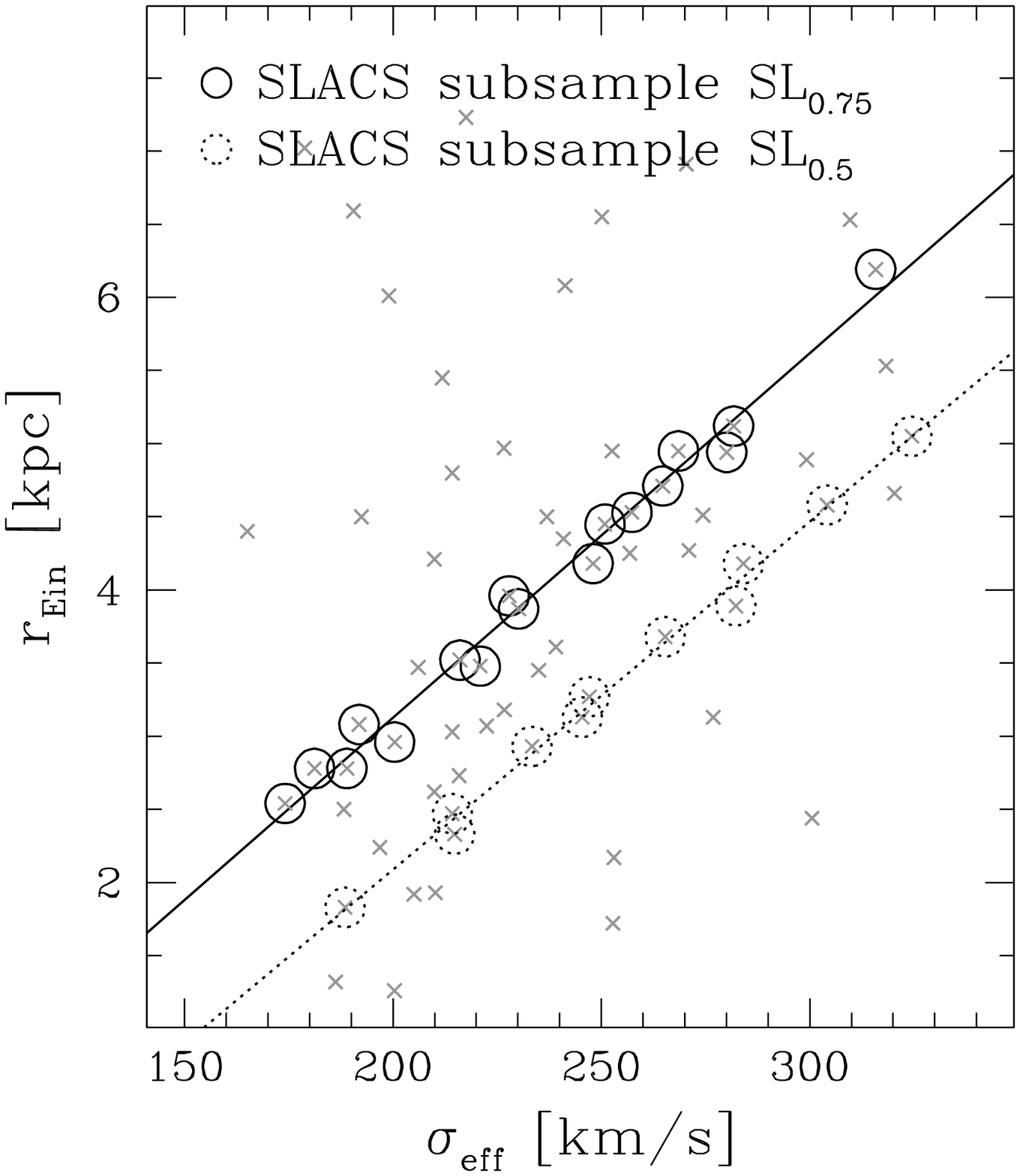}
\caption{Einstein radii $\rein$ and effective velocity
    dispersions $\siggal$ of SLACS galaxies from \citet{Aug09} are
    shown by the crosses. Dotted open circles highlight lenses that
    follow eq.~(\ref{def:subA}), shown by the dotted line, and form
    subsample $\sla$; solid open circles are for subsample $\slb$
    (i.e. galaxies that follow eq.~\ref{def:subB}, shown by the solid
    line).}
\label{fig:fitrelation}
\end{figure}

The gravitational potential of the galaxies is assumed to be
axisymmetric. Contrasting other methods like using Jeans equations,
the Schwarzschild technique allows the exploration of all possible
orbit configurations. The Coma sample is unique in being the only
larger sample of axisymmetric dynamical models including dark
matter. Previous modelling attempts either assumed spherical symmetry
\citep{Ger01} or did not include dark matter explicitly \citep{Cap06}.

\subsection{Stellar population models}
\label{subsec:gradients}
Stellar ages, metallicities, [$\alpha$/Fe] ratios and $R$-band stellar
mass-to-light ratios are determined by fitting the single stellar
population models of \citet{Mar98,Mar05} with $\alpha$-elements
overabundance of \citet{DTho03} to the Lick indices H$\beta$, $\langle
\mathrm{Fe} \rangle$, $[\mathrm{MgFe}]$ and Mg$\, b$.  Two
initial-stellar-mass functions are considered. Firstly, the Salpeter
IMF ($\mlsalp$, with mass limits of $0.1 \, \msun$ and $100 \, \msun$)
and, secondly, the Kroupa IMF ($\mlkroupa$, with the same mass limits,
but a shallower slope for stars below $0.5 \, \msun$).  The Salpeter
IMF implies more low-mass stars and a higher mass-to-light ratio.  In
the $R$-band the scaling between the two cases is $\mlsalp \approx
1.56 \, \mlkroupa$ \citep{Sal55,Kro01}. When the IMF does not need to
be specified, we will refer to stellar-population mass-to-light ratios
as $\mlssp$.

Stellar population properties are calculated at each radius with
observations. Salpeter $R$-band mass-to-light ratios are shown in
Fig.~\ref{fig:salpeter_data}.  The light-weighted averages within
$\reff$ are indicated by the dashed lines and form the basis for the
remainder of this paper (Tab.~\ref{tab:dmfrac}). For some galaxies
there are systematic differences between the two sides of a slit which
are slightly larger than the statistical errors (e.g. along the
major-axis of GMP0144). These systematic uncertainties are not
included in the errors quoted in Tab.~\ref{tab:dmfrac}.  Irrespective
of the metalicity gradients present in early-type galaxies
\citep{Meh03}, mass-to-light ratio gradients in the R-band are
generally small for the Coma galaxies
(cf. Fig.~\ref{fig:salpeter_data}). The results presented here do not
depend significantly on the averaging radius. Its choice is driven by
the most massive sample objects.  Their spectroscopic data reach only
out to $r_\mathrm{obs,max} \approx \reff$ and, for the purpose of
homogeneity, we restrict the averaging in other galaxies to the same
radius, even if the data extend further out.

For a subsample of the Coma galaxies studied here, multi-band
photometry from the Sloan Digital Sky Survey (SDSS; \citealt{Yor00})
is available. Fig.~\ref{fig:farben} compares observed SDSS colours
with the predictions of our best-fit SSP models. We have applied the
same colour corrections to the models as discussed in \citet{Sag10}
plus an additional $i-z = -0.05$ (Maraston, private communication), to
take into account the recent improvements in the calibration of the
Maraston SSP models \citep{Mar09}.  The vertical error-bars indicate
the 68\% confidence region derived from the observational
errors. Model colours are averaged inside $\reff$. They fit well to
the SDSS colours with average differences smaller than $\la 0.01 \,
\mathrm{mag}$.

\citet{Gri10} used observed SDSS colours to derive photometric stellar
population parameters for some of our Coma galaxies. In contrast to
the analysis presented here, they (1) assume a solar metallicity for
all galaxies (but see the middle panel of Fig.~\ref{fig:ratml_pop})
(2) allow for an extended star-formation history and (3) use the
\citet{Mar05} models without colour corrections.  In
Fig.~\ref{fig:grillo} we plot their photometric
$\mlkroupa^\mathrm{SDSS}$ (scaled to the Kroupa IMF) against our
$\mlkroupa$.  On average, both approaches yield consistent results
($\langle \mlkroupa/\mlkroupa^\mathrm{SDSS} \rangle = 1.11$), though
the rms-scatter ($\pm 0.35$) is large.

\section{Projected masses from dynamics and lensing}
\label{sec:projmass}
As it has been stated in Sec.~\ref{sec:intro}, the dynamical models
rely on the assumption of axial symmetry and steady state dynamics. To
check how accurately these assumptions are fulfilled in real galaxies
we first compare our dynamical models against gravitational lensing
results. The latter constrain the total projected mass inside a
cylinder delimited by the Einstein radius $\rein$ of the lens and are
less affected by symmetry assumptions \citep{Koc91}. 

\subsection{Lens selection}
\label{subsec:lensselect}
The Einstein radius of a gravitational lens results from two
independent properties of a lensing configuration. Firstly, from the
physical deflection angle that the lensing galaxy gives rise to
according to its gravity. It depends only on the mass distribution of
the foreground galaxy. Secondly, from projection factors that depend
on the distances of the foreground lens and the background source,
respectively. For the Coma galaxies we only know their mass
distributions, but they are not part of real lenses. To compare them
with observed gravitational lenses, we need to define an appropriate
lensing configuration for each Coma galaxy. Here we do this implicitly
by seeking for lensing galaxies that accidentally happen to fall on a
linear relation $\rein(\siggal)$ between the Einstein radius and the
effective velocity dispersion. Defining a fiducial Einstein radius for
each Coma galaxy according to the same $\rein(\siggal)$ then ensures,
that there is at least a subsample of real lenses with similar lens
configurations at a given $\siggal$. The Coma galaxies can be compared
to the lenses in such a subsample, but not to the rest of the lensing
galaxies. The form of the selection function is arbitrary, any other
function would serve equally well. We choose a linear relation for
simplicity.

Fig.~\ref{fig:fitrelation} shows the observed Einstein radii $\rein$
of SLACS lenses from \citet{Aug09} against their velocity
dispersions\footnote{Note that for Fig.~\ref{fig:fitrelation} we have
  used the correction of \citet{Cap06} to transform the measured SDSS
  aperture dispersions into $\siggal$.}. The large circles show two
different subsamples of lenses constructed as outlined above. The
lenses are selected to deviate by less than $0.2 \, \kpc$ from two
arbitrary linear selection functions $\rein(\siggal) = 0.025 \,
\siggal - 3.0$ and $\rein(\siggal) = 0.025 \, \siggal - 1.9$. After
having selected the lenses we fit a straight line to each subsample in
order to determine the actual best-fit $\rein(\siggal)$ that is used
to define fiducial Einstein radii for Coma galaxies. The fitted
relations differ only slightly from the original selection functions.
For subsample $\sla$ (dotted) we get
\begin{equation}
\label{def:subA}
\frac{\rein}{\kpc} = 0.0238 \times \frac{\siggal}{\kms} - 2.6709
\end{equation}
and for subsample $\slb$ (solid)
\begin{equation}
\label{def:subB}
\frac{\rein}{\kpc} = 0.0249 \times \frac{\siggal}{\kms} - 1.8564.
\end{equation}
The two subsamples are constructed as a compromise between (1) ending
up with a sizeable number of galaxies in each subsample and (2)
yielding sufficiently different subsamples to allow for a comparison
between Coma and lensing galaxies at different physical scales. For
subsample $\sla$ we get an average $\langle \rein/\reff \rangle
\approx 0.5$ (11 lenses), while for subsample $\slb$ it is
$\langle \rein/\reff \rangle \approx 0.75$ (17 lenses).

%%%%%%%%%%%%%%%%%%%%%%%%%%%%%%%%%%%%%%%%%%
% Ups_dyn/Ups_Krou vs mass
%%%%%%%%%%%%%%%%%%%%%%%%%%%%%%%%%%%%%%%%%%
\begin{figure*}\centering
\begin{minipage}{166mm}
\includegraphics[width=144mm,angle=0]{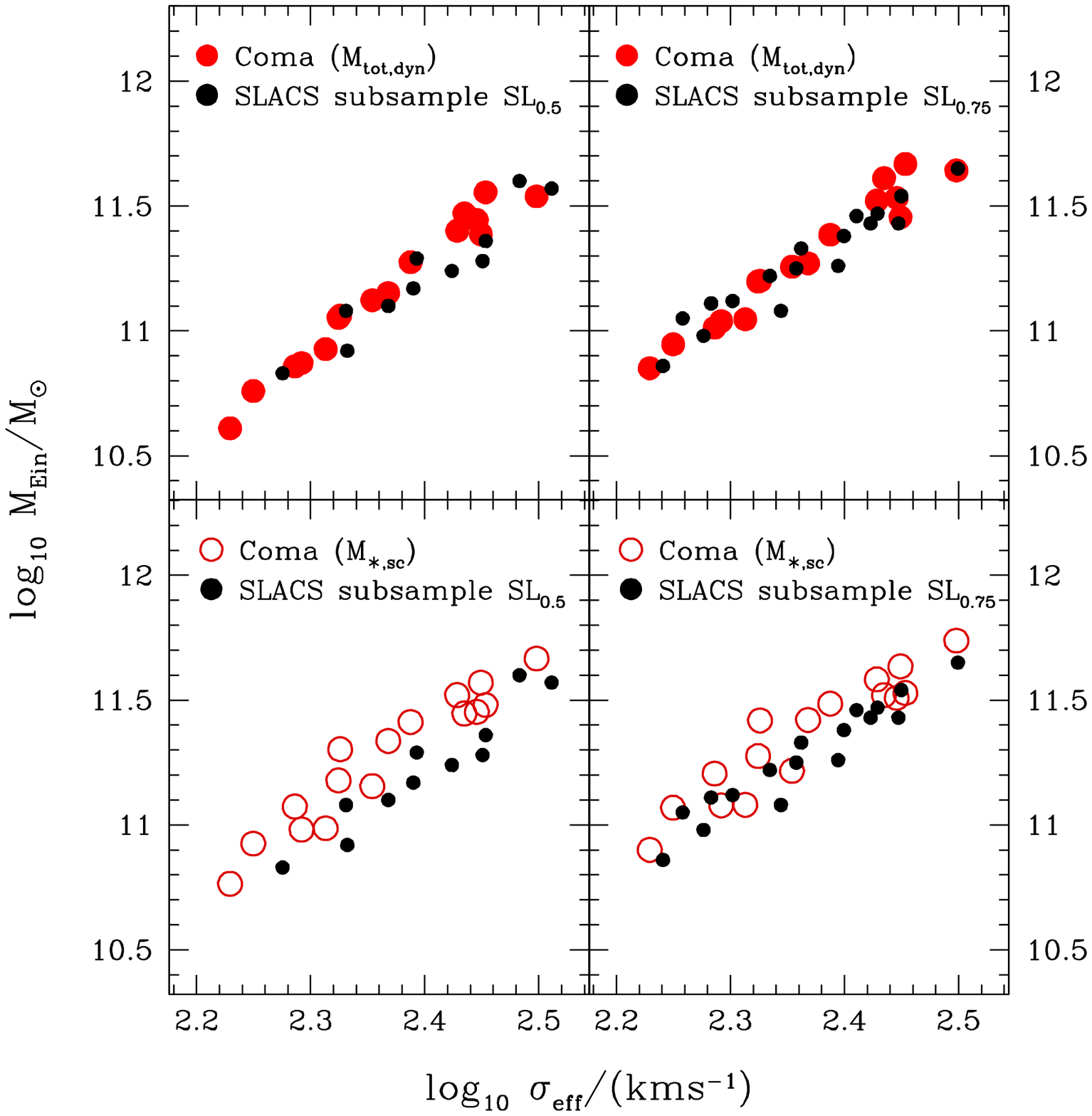}
\caption{The projected total (luminous+dark) mass $\mein$ within a
  fiducial Einstein radius $\rein$. Coma galaxies are indicated by the
  large symbols.  Top row: two-component models with dark matter halos
  ($\rho = \mldyn \times \nu + \rho_\mathrm{DM,dyn}$ in equation
  \ref{eq:meindef}); bottom row: dynamical mass under the assumption
  that mass follows light ($\rho = \mlsc \times \nu$).  Small circles:
  total projected masses of SLACS galaxies \citep{Aug09}. In the
  left-hand panels we compare Coma galaxies with the SLACS subsample $\sla$
  using fiducial Einstein radii calculated via eq.~(\ref{def:subA}). In
  the right-hand panels we compare to subsample $\slb$ (using
  eq.~\ref{def:subB}). Masses are plotted against the average velocity
  dispersion $\siggal$ inside $\reff$.}
\label{fig:projmass_tot}
\end{minipage}
\end{figure*}

%%%%%%%%%%%%%%%%%%%%%%%%%%%%%%%%%%%%%%%%%%
% Ups_dyn/Ups_Krou vs mass
%%%%%%%%%%%%%%%%%%%%%%%%%%%%%%%%%%%%%%%%%%
\begin{figure*}\centering
\begin{minipage}{166mm}
\includegraphics[width=144mm,angle=0]{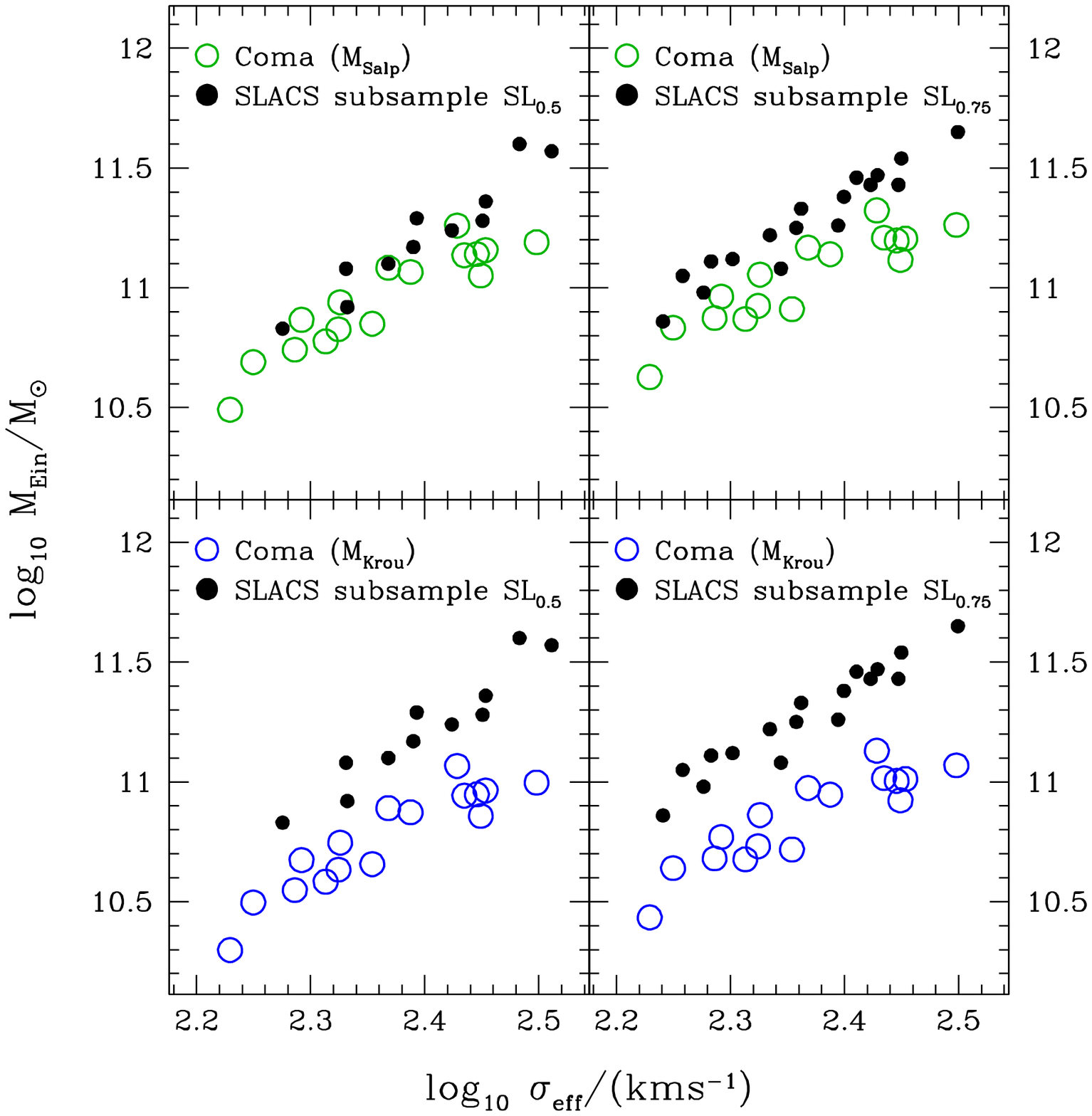}
\caption{As Fig.~\ref{fig:projmass_tot}, but for stellar population
  masses.  Top row: Salpeter IMF ($\rho = \mlsalp \times \nu$); bottom
  row: Kroupa IMF ($\rho = \mlkroupa \times \nu$).}
\label{fig:projmass_ssp}
\end{minipage}
\end{figure*}

\subsection{The total mass}
\label{subsec:totproj}
The large symbols in Fig.~\ref{fig:projmass_tot} show projected,
integrated masses of Coma galaxies and SLACS lenses as a function of
galaxy velocity dispersion $\siggal$.  Coma masses are calculated from
the integral
\begin{equation}
\label{eq:meindef}
\mein \equiv \int_{-10 \, \reff}^{10 \, \reff} \mathrm{d}z \int_0^{2\pi} \mathrm{d}\varphi \int_0^{\rein(\siggal)} \rho \, r \, \mathrm{d}r,
\end{equation}
where ($r,\varphi$) are polar coordinates on the sky and $z$ is the
direction of the line-of-sight. Formally, the integral in
eq.~(\ref{eq:meindef}) should be calculated over $-\infty \leq z \leq
+\infty$, but we limited it over $-10 \, r_{\rm eff} \leq z \leq +10 \,
r_{\rm eff}$. For an isothermal sphere this cut-off results in a
negligible underestimation of the integral ($\approx 4 \, \%$). Coma
galaxy velocity dispersions $\siggal$ are measured by (1)
reconstructing the line-of-sight velocity distributions (LOSVDs) from
the kinematic moments, (2) coadding all LOSVDs inside $\reff$, each
weighted by its projected light, and (3) fitting a Gaussian to the
resulting LOSVD. They are listed in Tab.~\ref{tab:dmfrac}.  Because of
their higher signal-to-noise, only the major-axis data have been
considered.  The upper integration limit $\rein(\siggal)$ in
eq.~(\ref{eq:meindef}) refers to eq.~(\ref{def:subA}) for the
comparison with SLACS subsample $\sla$ and to eq.~(\ref{def:subB}) for
the comparison with subsample $\slb$. In Fig.~\ref{fig:projmass_tot}
we only show those SLACS lenses that belong to either subsample $\sla$
(left-hand panels) or subsample $\slb$ (right-hand panels),
respectively.

The top row is for the total cylindrical mass of our standard
two-component models with dark matter halos (i.e. with the best-fit
density $\rho = \mldyn \times \nu + \rho_\mathrm{DM,dyn}$ in
eq.~\ref{eq:meindef}). The good agreement with the lensing results is
reassuring (the average mass offset is $0.05$ dex for subsample $\sla$
and $0.02$ dex for subsample $\slb$). It implies that the two
completely independent methods yield consistent results. Moreover, the
scatter in the dynamical masses is not larger than in the lensing
masses. Consequently, strong deviations from axisymmetry are unlikely
in the Coma galaxies. As shown in \citet{Tho07A}, strong triaxiality,
if not accounted for in the models, can bias dynamical masses by a
factor of up to two, depending on viewing angle.  Assuming random
viewing angles, strongly triaxial mass distributions would therefore
likely cause a significant scatter in the dynamical masses,
which is however not observed (but see \citealt{Ven09}).

The bottom row of Fig.~\ref{fig:projmass_tot} is for projected masses
of self-consistent dynamical models in which all the mass is assumed
to follow the light (i.e. $\rho = \mlsc \times \nu$). These models are
not consistent with the lensing masses. The discrepancy is larger for
the SLACS subsample $\sla$ than for subsample $\slb$ because the
smaller Einstein radii of subsample $\sla$ emphasise the central
regions in the comparison. As it has been shown earlier
(e.g. \citealt{Ger01}, \citealt{Tho07B}), the mass distribution in
early-type galaxies follows the light in the inner regions
($\rho_\mathrm{in} \approx \mldyn \times \nu_\mathrm{in}$), but has an
additional component in the outer parts ($\rho_\mathrm{out} \approx
\mldyn \times \nu_\mathrm{out} + \rhodm$).  Then, assuming that all
the mass follows the light requires $\mlsc > \mldyn$ to include the
outer dark matter ($\mlsc \times \nu_\mathrm{out} \approx
\rho_\mathrm{out} \approx \mldyn \times \nu_\mathrm{out} + \rhodm$).
However, the central regions become proportionally more massive, too,
such that $\mlsc \times \nu_\mathrm{in} > \mldyn \times
\nu_\mathrm{in} \approx \rho_\mathrm{in}$. This explains why the
offset in the lower-left panel of Fig.~\ref{fig:projmass_tot} is larger
than in the lower-right one.

\subsection{Stellar population masses}
In Fig.~\ref{fig:projmass_ssp} we show projected {\it stellar} masses
(i.e. using $\rho = \mlsalp \times \nu$ and $\rho = \mlkroupa \times
\nu$, respectively, in eq.~\ref{eq:meindef}).  The open dots in these
panels represent the projected {\it stellar} masses for either the
Salpeter IMF (top row) or the Kroupa IMF (bottom row). Kroupa stellar
masses are always below lensing masses. The mass difference between
the lenses and the Kroupa masses increases from subsample $\sla$ to
subsample $\slb$ for the same reason discussed at the end of 
Sec.~\ref{subsec:totproj}. It could be due to dark matter. Salpeter
stellar masses are likewise consistent with the lensing results. In
low-dispersion galaxies the IMF cannot be much steeper than Salpeter
as otherwise the implied stellar masses would exceed the total
observed lens masses. In high-dispersion galaxies the Salpeter stellar
masses are however not enough to explain the total lensing
masses. Then, if all the lensing mass was stellar, the IMF would have
to change with galaxy velocity dispersion and in massive early-types
the stellar mass per stellar light would have to be larger than for a
Salpeter IMF (or any equivalent top-heavy IMF). If the IMF is
constant, then the top-left panel of Fig.~\ref{fig:projmass_ssp}
provides direct evidence for the presence of dark matter in
high-dispersion early-type galaxies.

%%%%%%%%%%%%%%%%%%%%%%%%%%%%%%%%%%%%%%%%%%
% Ups_dyn/Ups_Krou vs sigma
%%%%%%%%%%%%%%%%%%%%%%%%%%%%%%%%%%%%%%%%%%
\begin{figure}\centering
\includegraphics[width=84mm,angle=0]{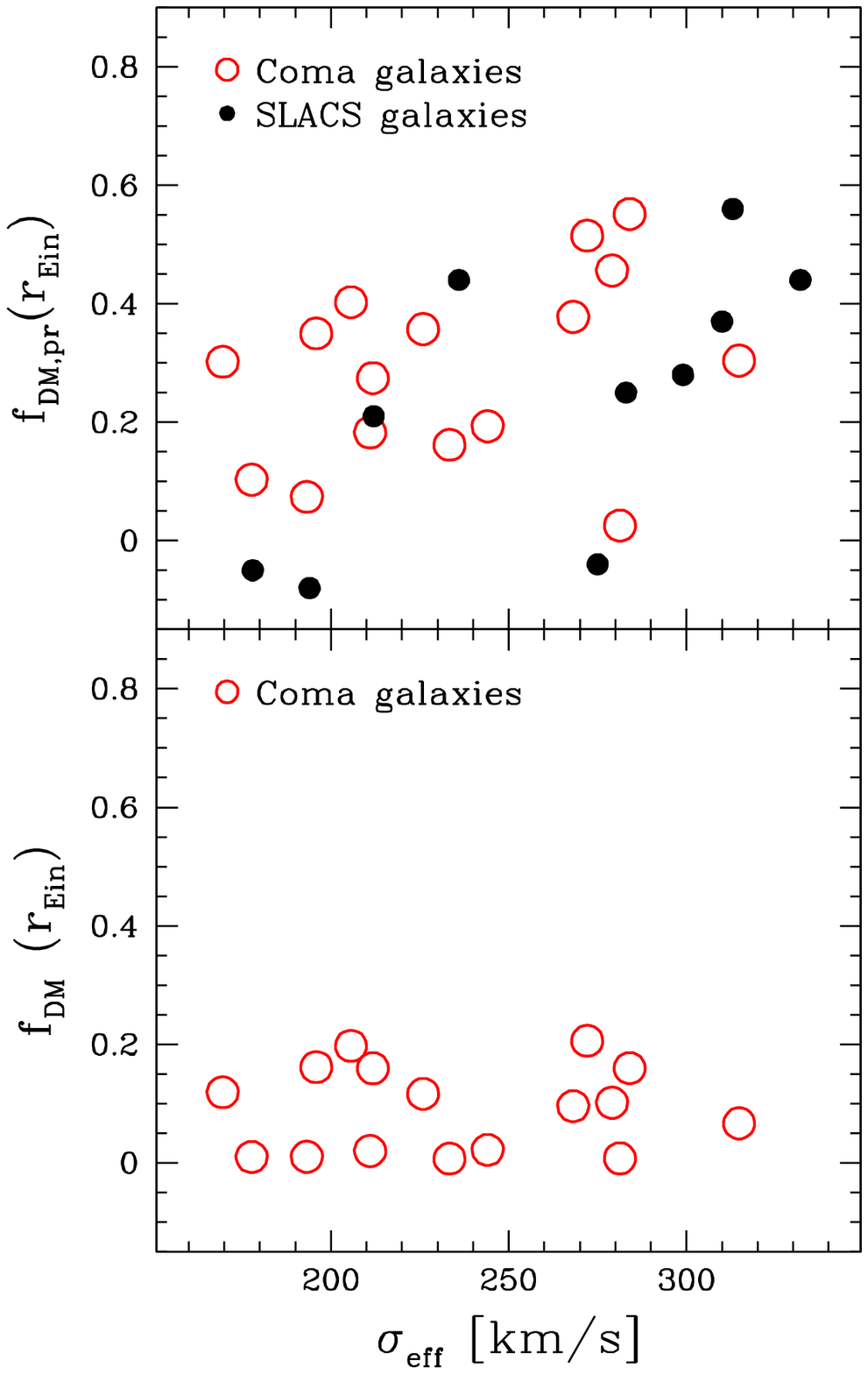}
\caption{Dark matter fractions inside the Einstein radius.
  Top: projected dark matter fraction; bottom: deprojected dark matter
  fraction.  Red/open circles: Coma galaxies of this work; dark/filled
  circles in the top panel: SLACS galaxies from \citet{Koo06}. Coma
  galaxy dark matter fractions are plotted against the velocity
  dispersion $\siggal$ inside $\reff$. SLACS galaxies are plotted
  against the measured velocity dispersion inside the SDSS
  aperture.}
\label{fig:projdmfrac}
\end{figure}

\subsection{Luminous and dark matter separated}
In our standard two-component models that take into account the
detailed, radially resolved stellar kinematics, the large projected
masses of high-dispersion galaxies do not entirely originate from
luminous mass. This is shown in the top panel of
Fig.~\ref{fig:projdmfrac} (projected dark matter fractions inside the
Einstein radius). Note that here we compare to the results of
\citet{Koo06}, who provided a combined lensing and dynamics analysis
of the first SLACS galaxies. We applied a similar lens selection as
described in Sec.~\ref{subsec:lensselect} (using $\rein = 0.03244
\times \siggal - 4.6324$). Our dynamically derived dark matter
fractions are in good agreement with those from \citet{Koo06}. As
already stated above, they increase with $\siggal$. For comparison, we
have also plotted the corresponding deprojected dark matter fractions
(cf. eq.~\ref{eq:dmfrac}) inside the same three dimensional radius
$\rein(\siggal)$ in the bottom panel of
Fig.~\ref{fig:projdmfrac}. They are generally lower and do not vary
with $\siggal$. Projection effects therefore contribute to the trends
in Figs.~\ref{fig:projmass_tot}-\ref{fig:projdmfrac}. To get a better
understanding about the stellar IMF and dark matter distribution it is
necessary to analyse the intrinsic three dimensional properties of the
galaxies. This will be done in the following Secs.~\ref{sec:ml} and
\ref{sec:dm}.

\section{Dynamical luminous mass versus stellar population mass}
\label{sec:ml}
Fig.~\ref{fig:ml_sig} shows dynamical and stellar-population
mass-to-light ratios versus $\siggal$. The dynamical $\mldyn$ exhibit
a broad distribution ranging from $\mldyn \approx 2$ to $\mldyn
\approx 10$. In contrast, for the majority of Coma galaxies the
stellar population $\mlssp$ are almost constant with $\mlsalp \approx
5-6$ for the Salpeter IMF and $\mlkroupa \approx 3-4$ for the Kroupa
IMF. In addition, while the dynamical $\mldyn$ clearly increase with
galaxy velocity dispersion, there is no similar correlation between
$\siggal$ and stellar-population $\mlssp$.

There are two galaxies (GMP0756 and GMP1176) with distinctly lower
stellar mass-to-light ratios than in the rest of the sample. As
Fig.~\ref{fig:ml_tau} shows, these two galaxies have overall young
stellar populations. In general, one can read off from
Figs.~\ref{fig:ml_sig} and \ref{fig:ml_tau} that dynamical $\mldyn$
depend on $\siggal$ but not primarily on the stellar population age,
while $\mlssp$ mostly reflect stellar ages and do not show any
dependency on $\siggal$.

\subsection{Variation in the stellar IMF?}
\begin{figure*}\centering
\begin{minipage}{166mm}
\includegraphics[width=144mm,angle=0]{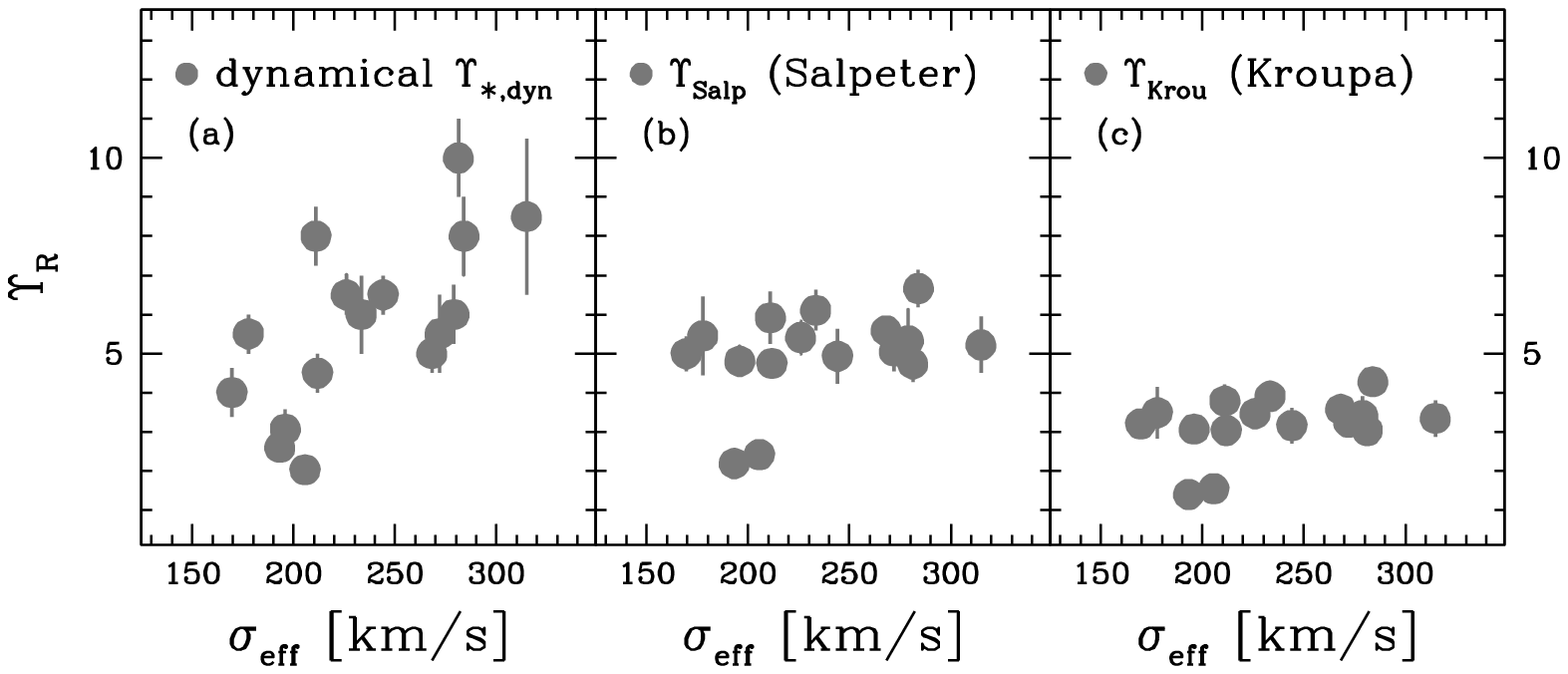}
\caption{Stellar mass-to-light ratios (R-band) against average
  velocity dispersion $\siggal$ inside $\reff$.  From left-to-right:
  a) dynamical $\mldyn$, b) stellar-population $\mlsalp$ for the
  Salpeter IMF and c) stellar-population $\mlkroupa$ for the Kroupa
  IMF. Circles: E/S0 and S0 galaxies; squares: ellipticals.}
\label{fig:ml_sig}
\end{minipage}
\end{figure*}

\begin{figure*}\centering
\begin{minipage}{166mm}
\includegraphics[width=144mm,angle=0]{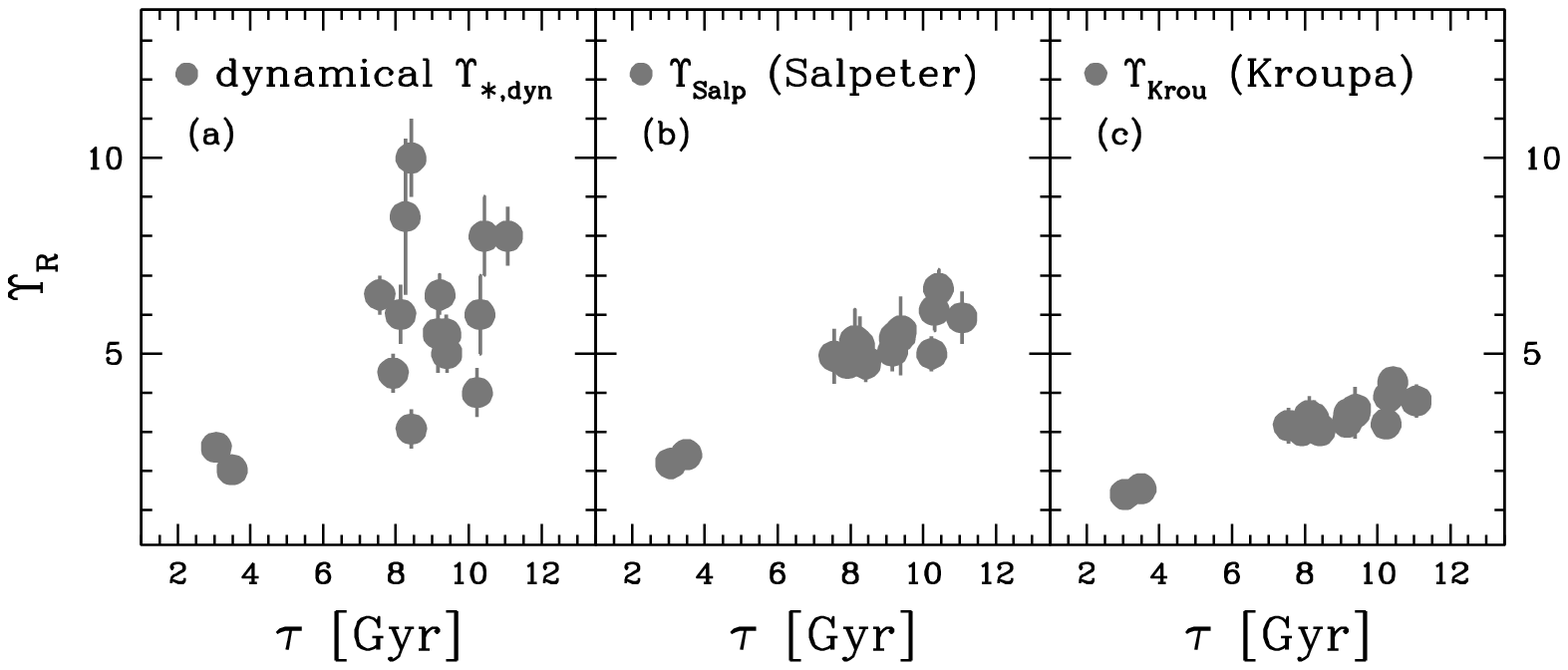}
\caption{As Fig.~\ref{fig:ml_sig}, but stellar mass-to-light ratios
  are plotted against stellar population age $\tau$.}
\label{fig:ml_tau}
\end{minipage}
\end{figure*}

\begin{figure}\centering
\includegraphics[width=84mm,angle=0]{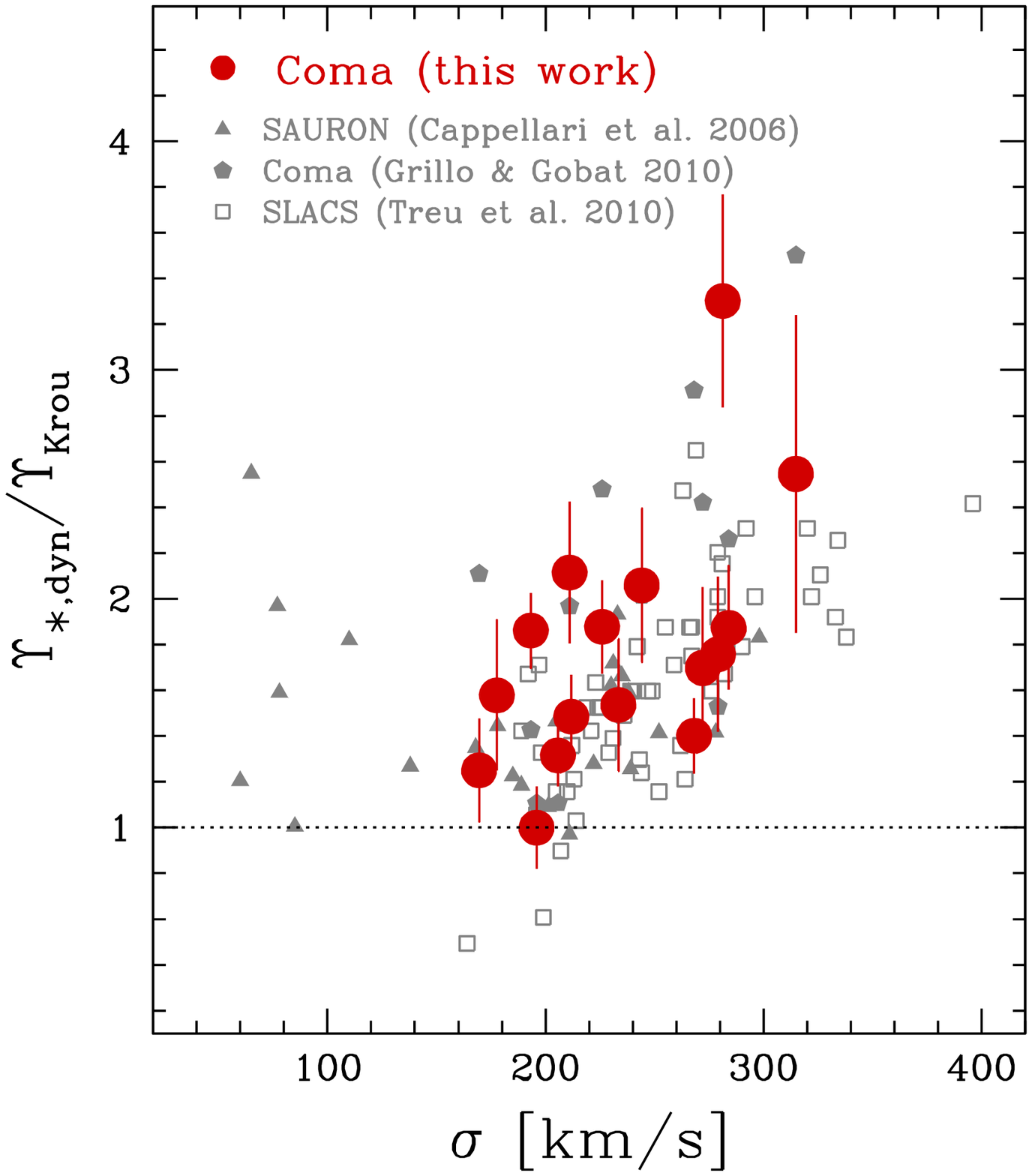}
\caption{Ratio $\ratml$ versus velocity dispersion $\sigma$.  Large,
  filled: spectroscopic $\mlkroupa$ and orbit-based dynamical models
  accounting for dark matter (this work); triangles: spectroscopic
  $\mlkroupa$ and orbit-based dynamical models neglecting dark matter
  \citep{Cap06}; pentagons: photometric $\mlkroupa$ \citep{Gri10} and
  orbit-based dynamical models accounting for dark matter; open
  squares: photometric $\mlkroupa$ and combined lensing + dynamics
  models \citep{Tre10}. For all but the SLACS galaxies the average
  velocity dispersion $\siggal$ inside $\reff$ is plotted. The SLACS
  dispersions are the average over the spectroscopic aperture of the
  SDSS survey.}
\label{fig:comp}
\end{figure}

Since dynamical and stellar population masses scale differently with
galaxy velocity dispersion (cf. Fig.~\ref{fig:ml_sig}), the ratio
$\mldyn/\mlssp$ has to vary with galaxy $\siggal$ (for any fixed IMF).
This is explicitly shown by the large/red symbols in
Fig.~\ref{fig:comp}.  In addition to our results, the figure also
combines work from other groups.

Pentagons represent the stellar-population analysis of a subsample of
our Coma galaxies by \citet{Gri10}
(cf. Sec.~\ref{subsec:gradients}). The pentagons differ from the large
filled symbols only in terms of $\mlkroupa$. The velocity dispersions
and dynamical $\mldyn$ are the same.

Triangles are for the SAURON survey \citep{Cap06}. In terms of both,
the stellar population analysis (based on spectral absorption line
indices) as well as the dynamical modelling (orbit-based), they can be
most directly compared to the Coma galaxies of this work. Note,
however, that \citet{Cap06} measured only the sum of luminous and dark
mass, assuming the latter to contribute only a small amount of mass in
the central galaxy regions observed with SAURON ($r_\mathrm{obs,max}
\la \reff$). Adopting equivalent modelling assumptions for the Coma
galaxies yields mass-to-light ratios typically $10-20 \, \%$ higher
than compared with stellar $\mldyn$ from models where dark matter is
accounted for explicitly (cf. Tab.~\ref{tab:dmfrac}).  The overall
distributions of $\ratml$ are nevertheless similar in both samples.

%%%%%%%%%%%%%%%%%%%%%%%%%%%%%%%%%%%%%%%%%%
% M/L vs mass
%%%%%%%%%%%%%%%%%%%%%%%%%%%%%%%%%%%%%%%%%%

%%%%%%%%%%%%%%%%%%%%%%%%%%%%%%%%%%%%%%%%%%
% M/L vs mass
%%%%%%%%%%%%%%%%%%%%%%%%%%%%%%%%%%%%%%%%%%

Finally, Fig.~\ref{fig:comp} also includes SLACS galaxies, analysed
with a combined dynamics and lensing approach \citep{Tre10}. For these
galaxies, the ratio of the total versus the stellar mass (the latter
derived from broad-band colours) inside the Einstein radius (typically
of the order of $\reff/2$) is plotted along the y-axis.

In all the samples included in Fig.~\ref{fig:comp} dynamical (or
lensing) stellar masses systematically exceed the required masses for
a Kroupa IMF.  Above $\siggal \ga 150 \, \kms$ the ratio
$\mldyn/\mlkroupa$ tends to increase with velocity dispersion. For
lower mass galaxies the ratio becomes uncertain due to the more
frequent presence of multiple stellar populations
(e.g. \citealt{Cap06}). Moreover, for low-mass galaxies the assumption
that all the mass follows the light is in conflict with gravitational
lensing masses (cf. Sec.~\ref{sec:projmass}).

If the increase of $\mldyn/\mlkroupa$ with $\siggal$ was a pure
stellar population effect then we would have to assume that the IMF is
not universal. Interpreted in this way, Fig.~\ref{fig:comp} would
imply the IMF in high-dispersion galaxies to produce either more
low-mass stars than the Kroupa IMF (i.e. being Salpeter-like) or more
stellar remnants (i.e. being top-heavy). A higher fraction of low-mass
stars would reduce the number of SNe of type II (per stellar mass) and
would lead to an overall lower metallicity. Thus, if the IMF changed
from Kroupa towards Salpeter, then the increase in $\mldyn/\mlkroupa$
would be expected to come along with a decrease in [Z/H]. Instead, a
top-heavy IMF enhances the importance of type II SNe over type Ia
SNe. Accordingly, if the IMF changed from Kroupa towards being
top-heavy then one would expect higher [$\alpha$/Fe] in galaxies with
higher $\mldyn/\mlkroupa$ \citep{Tho99,Gra10}.

%%%%%%%%%%%%%%%%%%%%%%%%%%%%%%%%%%%%%%%%%%
% Ups_dyn/Ups_Krou vs sigma
%%%%%%%%%%%%%%%%%%%%%%%%%%%%%%%%%%%%%%%%%%

\begin{figure*}\centering
\begin{minipage}{166mm}
\includegraphics[width=144mm,angle=0]{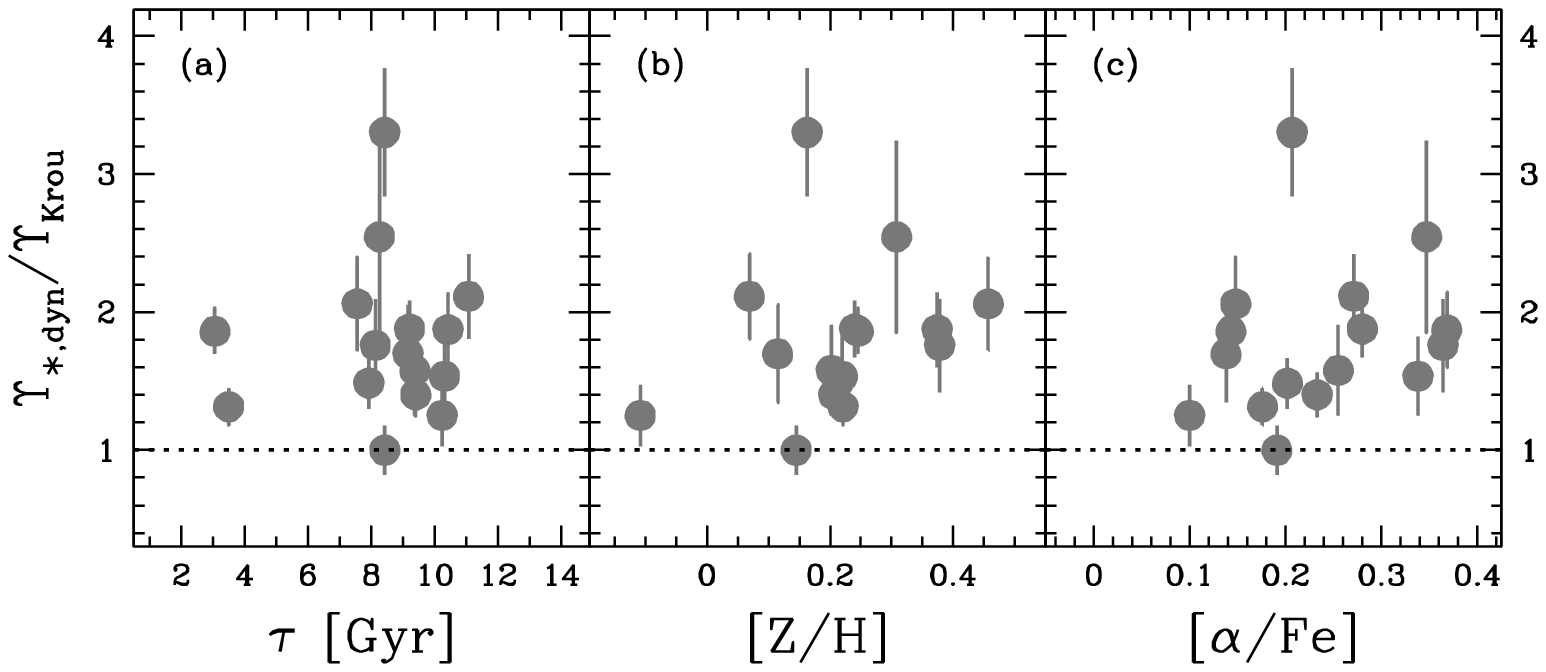}
\caption{Ratio $\mldyn/\mlkroupa$ against stellar population age
  $\tau$ (panel a), metallicity [Z/H] (panel b) and abundance ratio
  [$\alpha$/Fe] (panel c).}
\label{fig:ratml_pop}
\end{minipage}
\end{figure*}

Fig.~\ref{fig:ratml_pop} shows $\mldyn/\mlkroupa$ against stellar
population age $\tau$, metallicity [Z/H] and [$\alpha$/Fe]
ratio. There is no correlation with any stellar population parameter.
Note, however, that stellar metallicities and [$\alpha$/Fe] ratios do
not only depend on the stellar IMF, but also on the duration of the
star-formation episode(s), the depth of the galaxy potential well and
on evolutionary processes related to the cluster environment. In this
respect, the lack of evidence for an IMF change cannot be taken as a
proof for a constant IMF. Alternatives to IMF variation are discussed
in Secs.~\ref{subsec:mldynerr} and \ref{subsec:lightdm}.

\subsection{The shape of the stellar IMF}
\label{subsec:imfnorm}
\begin{figure}\centering
\includegraphics[width=84mm,angle=0]{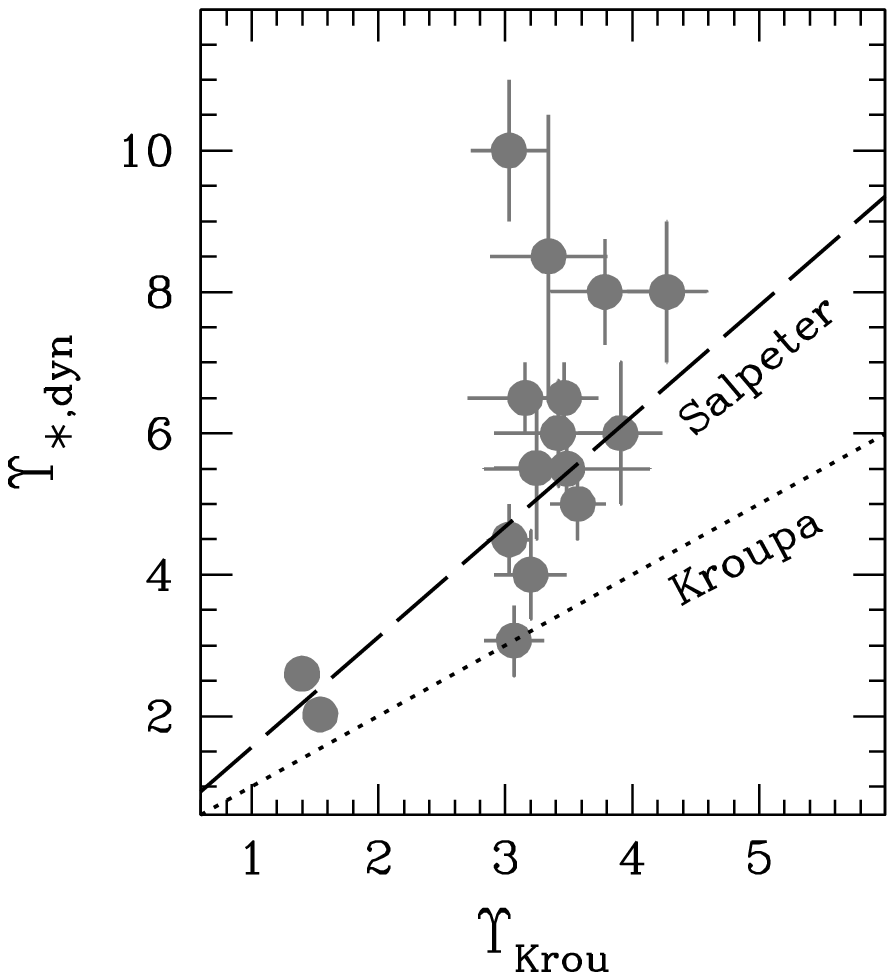}
\caption{Dynamical $\mldyn$ versus stellar-population
    mass-to-light ratio $\mlkroupa$. The dotted line shows the
    one-to-one relation for the Kroupa IMF. Since $\mlsalp = 1.56
    \times \mlkroupa$, the corresponding one-to-one relation for the
    Salpeter IMF occurs at larger $\mldyn$ (dashed line).}
\label{fig:ml_ml_krou}
\end{figure}

A direct comparison between dynamical and stellar-population
mass-to-light ratios is provided by Fig.~\ref{fig:ml_ml_krou}. As
already clear from the different scalings of $\mldyn$ on the one side
and $\mlssp$ on the other (cf. Figs.~\ref{fig:ml_sig} and
\ref{fig:ml_tau}), neither the Kroupa IMF nor the Salpeter IMF yields
a close match between $\mldyn$ and $\mlssp$.

Above the dotted line in Fig.~\ref{fig:ml_ml_krou} the luminous
dynamical mass is larger than the stellar mass required for the Kroupa
IMF, while below the line the dynamical $\mldyn$ is formally
insufficient for the Kroupa IMF. For the Salpeter IMF the
corresponding limit is shifted towards $\mldyn$ which are a factor
$1.6$ higher (dashed line).  Accordingly, all Coma galaxies are
compatible with a Kroupa IMF. The majority of the galaxies is also
consistent with a Salpeter IMF, but there is at least one galaxy for
which the dynamical $\mldyn$ is significantly lower than $\mlsalp$ (at
about the $3 \, \sigma$ level; GMP5975).  Concerning the total sample,
however, the Salpeter IMF fits the dynamical masses better than the
Kroupa IMF. The corresponding sample averages are $\langle
\mldyn/\mlsalp \rangle = 1.15$ for the Salpeter IMF (with an rms
scatter of $0.35$) and $\langle \mldyn/\mlkroupa \rangle = 1.8$ for
the Kroupa IMF, respectively.

%%%%%%%%%%%%%%%%%%%%%%%%%%%%%%%%%%%%%%%%%%
% Ups_dyn/Ups_Krou vs mass
%%%%%%%%%%%%%%%%%%%%%%%%%%%%%%%%%%%%%%%%%%

These conclusions also hold for other galaxy samples. Since $\mlsalp
\approx 1.6 \times \mlkroupa$, the one-to-one line for the Salpeter
IMF in Fig.~\ref{fig:comp} would occur at $\mldyn/\mlssp \approx
1.6$. Then, the Salpeter IMF provides {\it on average} a better match
with dynamical/lensing masses. In line with this, recent near-infrared
spectroscopic observations point towards a bottom-heavy IMF in massive
early-type galaxies as well \citep{Dok10,Dok11}. However, our
dynamical models as well as previous lensing studies
\citep{Fer10,Tre10} indicate that around $\siggal \la 200 \, \kms$ the
Salpeter stellar masses exceed the observed dynamical and/or lensing
limits. This rules out a Salpeter IMF for low-mass galaxies.

The results for the Coma galaxies are largely independent of the
parameterisation chosen for the dark matter halos. Fits with
logarithmic halos alone yield $\langle \mldyn^\mathrm{LOG}/\mlsalp
\rangle = 1.16$, while NFW halos result in $\langle
\mldyn^\mathrm{NFW}/\mlsalp \rangle = 1.06$. Both are consistent
within the rms scatter (about $\approx 0.32$).

\subsection{Uncertainties in population $\mlssp$}
Gas emission can refill the H$\beta$ line and lead to an overestimate
of stellar population ages and, then, of $\mlssp$. A young stellar
subpopulation (dominating in terms of light, but not in mass) can
likewise bias stellar ages and $\mlssp$, yet towards too low
values. In any case, $\mldyn/\mlssp$ would systematically decrease
with stellar population age. Fig.~\ref{fig:ratml_pop}a shows however,
that this is not the case in the Coma galaxies, such that a strong
bias due to gas emission or young stellar subpopulations is unlikely.

The stellar population parameters of the Coma galaxies are derived
from spectral indices and, thus, represent averages along the
line-of-sight. In other words, at a given radius of observation
$r_\mathrm{obs}$, the SSP parameters combine the properties of stars
with $r>r_\mathrm{obs}$. In contrast, dynamical models are most
sensitive to the mass distribution inside $r_\mathrm{obs}$. Projection
effects can therefore introduce a systematic bias between dynamical
$\mldyn$ and stellar population $\mlssp$ if the stellar-population
changes with radius.  A monotonic stellar population gradient is
enhanced after projection along the line-of-sight but diminished in
the cumulative mass-to-light ratio constrained by dynamical models
\citep{Tho06}. More specifically, a radial increase of $\Upsilon$
leads to an underestimation of $\mldyn/\mlssp$, a radial decrease of
$\Upsilon$ to an overestimation of $\mldyn/\mlssp$. For a rather steep
gradient of $\mathrm{d} \log \Upsilon/ \mathrm{d} \log r = \pm 0.23$
(a change by a factor of 1.7 per decade in radius), the expected
systematic difference between $\mldyn$ and $\mlssp$ would amount to
$\mp 30$ percent inside $\reff$ \citep{Tho06}. The observed gradients
in the Coma galaxies are however much smaller
(cf. Fig.~\ref{fig:salpeter_data}). Therefore systematics due to
projection effects seem negligible.

%%%%%%%%%%%%%%%%%%%%%%%%%%%%%%%%%%%%%%%%%%
% Ups_dyn/Ups_Krou vs sigma
%%%%%%%%%%%%%%%%%%%%%%%%%%%%%%%%%%%%%%%%%%

\subsection{Uncertainties and interpretation of dynamical $\mldyn$}
\label{subsec:mldynerr}
The dynamical mass-to-light ratios $\mldyn$ could be affected by
systematic biases in the modelling process, e.g. arising from false
symmetry assumptions.  As it has been stated in Sec.~\ref{sec:intro}
the luminous $\mldyn$ can be biased by a factor of up to two, if the
studied galaxies deviate significantly from the symmetry assumed in
our models \citep{Tho07A}. However, there is no evidence for the Coma
galaxies to be strongly non-axisymmetric. Firstly, they do not show
significant isophotal twists as would be indicative for
triaxiality. Secondly, as already discussed in
Sec.~\ref{sec:projmass}, the good match between dynamical and
strong-lensing masses provides further evidence that the obtained
dynamical masses are unbiased.

Even if the luminous $\mldyn$ are accurate, they might not represent
the galaxy {\it stellar} mass in a one-to-one fashion. Ambiguities can
come from any non-stellar mass that follows the light and contributes
to $\mldyn$.

Such a mass component could be gas loss during stellar evolution. It
is not included in our $\mlssp$, which only encompass the baryonic
mass locked in stars or stellar remnants.  For a 10 Gyr old population
the lost gas mass amounts to about 40 percent of the originally formed
stellar mass (e.g. \citealt{Mar05}).  Provided that it remains in the
galaxies and provided it follows the light distribution, it would
contribute to the dynamical mass $\mldyn$ and the actual baryonic mass
in stars and stellar remnants would only be $(\mlcor) \times L$ (with
$f_\ast \approx 0.6$). Correcting the dynamical $\mldyn$ for stellar
mass loss yields formally a good agreement with the Kroupa IMF:
$\langle (\mlcor)/\mlkroupa \rangle = 1.04 \pm 0.32$.

Theoretical arguments indicate that most of the lost gas is either
expelled from the galaxies or recycled into new stars. Only a few
percent of the original stellar mass is expected to remain in hot gas
halos around the galaxies (e.g. \citealt{Cio91,Dav91}). Fittingly,
observed gas mass fractions of hot X-ray halos around massive
early-type galaxies are typically less than a percent of the present
stellar mass (e.g. \citealt{Mat01}), such that stellar mass loss is
unlikely to explain the excess $\mldyn > \mlkroupa$.

A possible component of non-baryonic matter that follows the light is
discussed below in Sec.~\ref{subsec:lightdm}.

%%%%%%%%%%%%%%%%%%%%%%%%%%%%%%%%%%%%%%%%%%
% f_DM vs r_eff
%%%%%%%%%%%%%%%%%%%%%%%%%%%%%%%%%%%%%%%%%%
\begin{figure}\centering
\includegraphics[width=84mm,angle=0]{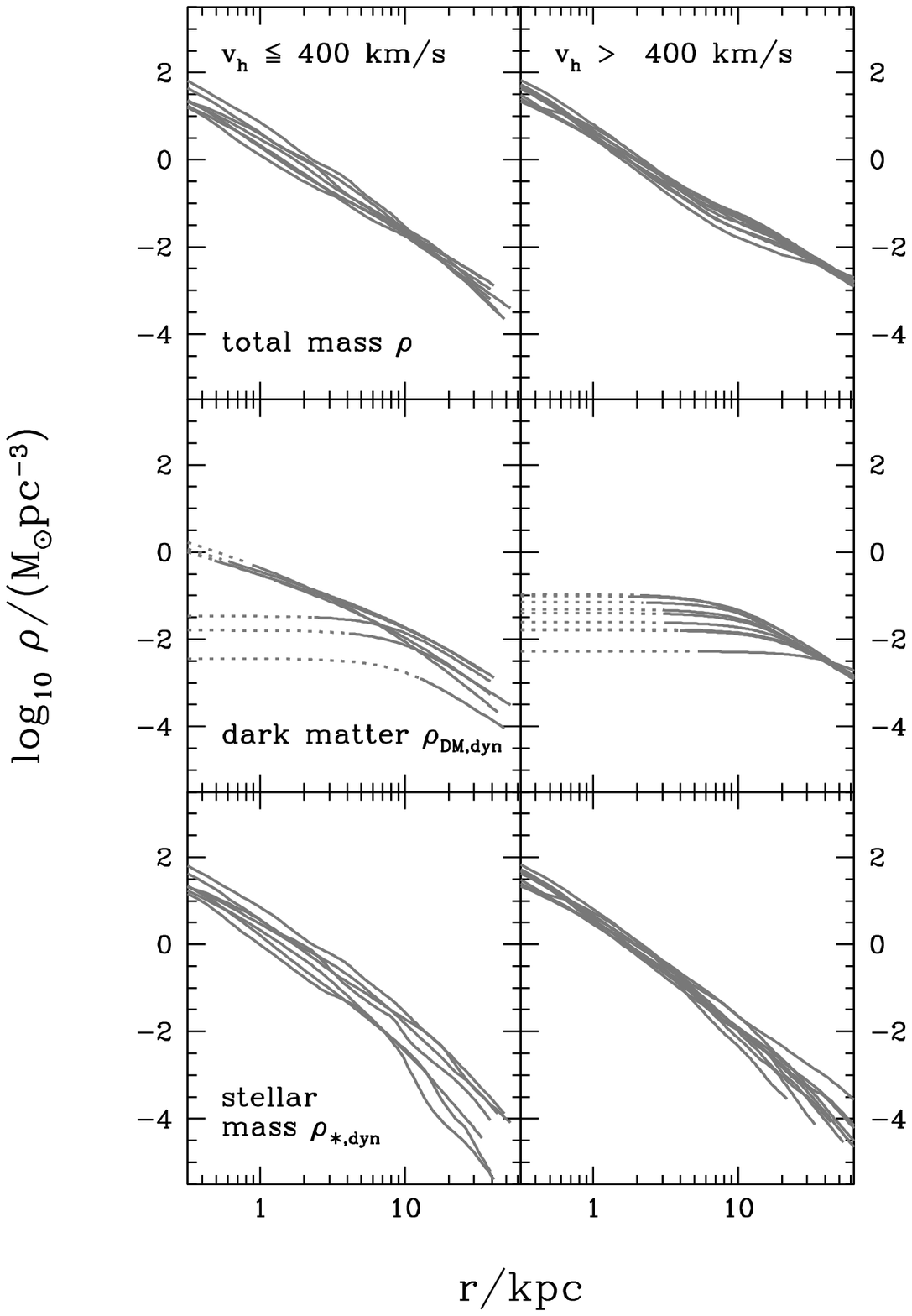}
\caption{From top to bottom: total (luminous + dark) three dimensional
  mass density $\rho$, dark matter density $\rho_\mathrm{DM,dyn}$ and
  luminous mass density $\rho_\mathrm{\ast,dyn}$.  All densities are
  spherically averaged. Left-hand panels: galaxies with less massive
  halos ($v_h \le 400 \, \kms$); right-hand panels: galaxies with
  massive halos ($v_h > 400 \, \kms$. In the middle row the dotted
  lines indicate the spatial region where $\rho_\mathrm{DM,dyn} <
  \rho_\mathrm{\ast,dyn}/10$.  }
\label{fig:densplot}
\end{figure}

\section{Dark matter}
\label{sec:dm}

\subsection{Mass that does not follow the light}

%%%%%%%%%%%%%%%%%%%%%%%%%%%%%%%%%%%%%%%%%%
% Ups_dyn/Ups_Krou vs mass
%%%%%%%%%%%%%%%%%%%%%%%%%%%%%%%%%%%%%%%%%%
\begin{figure*}\centering
\begin{minipage}{166mm}
\includegraphics[width=144mm,angle=0]{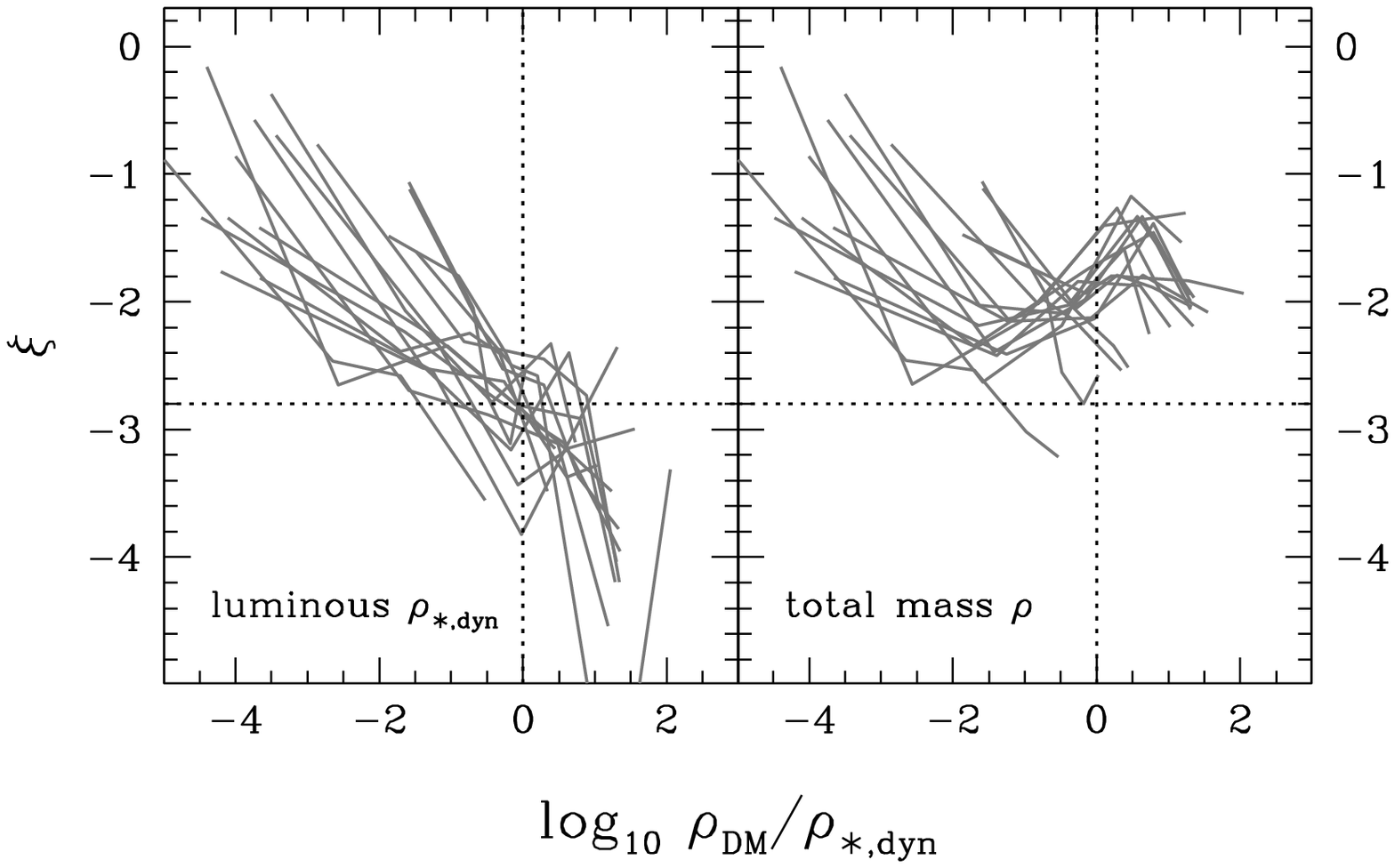}
\caption{Comparison of the local logarithmic density slope $\xi =
  \mathrm{d} \ln \rho(r)/\mathrm{d} \ln(r)$ with the density ratio
  $\rho_\mathrm{DM}(r)/\rho_\mathrm{\ast,dyn}(r)$ of dark to luminous
  matter at the same radius. Left panel: logarithmic slope of the
  luminous mass density $\rho_\mathrm{\ast,dyn} = \mldyn \times \nu$; right panel:
  logarithmic slope of the total mass density $\rho = \mldyn \times
  \nu + \rhodm$. Vertical dotted lines indicate where luminous and
  dark matter equalise. To the left of these lines luminous mass
  dominates (inner regions of the galaxies), to the right dark matter
  dominates (outer regions of galaxies). Horizontal lines are for $\xi
  = -2.8$, the luminous slope at which dark matter takes over luminous
  matter.}
\label{fig:slopes}
\end{minipage}
\end{figure*}

Fig.~\ref{fig:densplot} shows the spherically averaged three
dimensional density distributions of luminous and dark matter, as well
as the sum of both. The sample is subdivided into galaxies with
LOG-halo circular velocities $v_h \le 400 \, \kms$ and galaxies with
$v_h > 400 \, \kms$. The reason is that the latter galaxies have very
uniform outer dark and total mass density profiles. The corresponding
dark matter fractions (inside $\reff$) scatter around a mean of
$\langle \dmfrac \rangle = 23 \pm 17 \, \%$ and do not depend on
$\siggal$. Note that in the central galaxy regions our models
implicitly maximise the mass contribution from the light. Similar
maximum-bulge models for lensing galaxies yield dark matter fractions
around $25 \, \%$ as well \citep{Bar09}.

Fig.~\ref{fig:slopes} shows the logarithmic slope of the luminosity
density and the total mass density in the Coma galaxies. The slope is
plotted against the ratio of dark to luminous matter densities at the
same radii.  In the inner regions the slope of the total mass density
follows the light, since dark matter is negligible
($\rho_\mathrm{\ast,dyn} \gg \rhodm$), in the outer regions the total
mass density profile is flatter than the light profile. Overall, the
total mass density is roughly isothermal: $\rho \sim r^{-2}$. This
reflects the nearly flat circular velocity curves of early-type
galaxies \citep{Ger01,Tho07B}. Similar slopes for the total mass
distribution have been seen in lensing galaxies
\citep{Koo06,Bar11}. Fig.~\ref{fig:slopes} illustrates that the actual need
for a dark matter component in our models comes from the fact that the
outer mass distribution does not follow the light in early-type
galaxies. The radius where the density of dark matter takes over
luminous matter is roughly where the slope of the luminosity density
falls below $\xi \approx -2.8$ (indicated by the horizontal dotted
lines in Fig.~\ref{fig:slopes}).

\subsection{A component of dark matter that follows the light?}
\label{subsec:lightdm}
A galaxy might have more dark matter than captured by $\rhodm$ if some
fraction of the halo mass follows the light so closely that it is
mapped onto $\mldyn$ rather than $\rhodm$. In particular,
Fig.~\ref{fig:slopes} leaves the possibility open that this could
happen in the inner galaxy regions where the slope of the luminosity
distribution is $-2 \la \xi \la -1$. If the fraction of dark matter
that follows the light is larger in galaxies with higher $\siggal$,
then this would be a possible explanation for the trend between
$\mldyn/\mlkroupa$ and $\siggal$ seen in Fig.~\ref{fig:comp}.

%%%%%%%%%%%%%%%%%%%%%%%%%%%%%%%%%%%%%%%%%%
% Ups_dyn/Ups_Krou vs mass
%%%%%%%%%%%%%%%%%%%%%%%%%%%%%%%%%%%%%%%%%%
\begin{figure*}\centering
\begin{minipage}{166mm}
\includegraphics[width=144mm,angle=0]{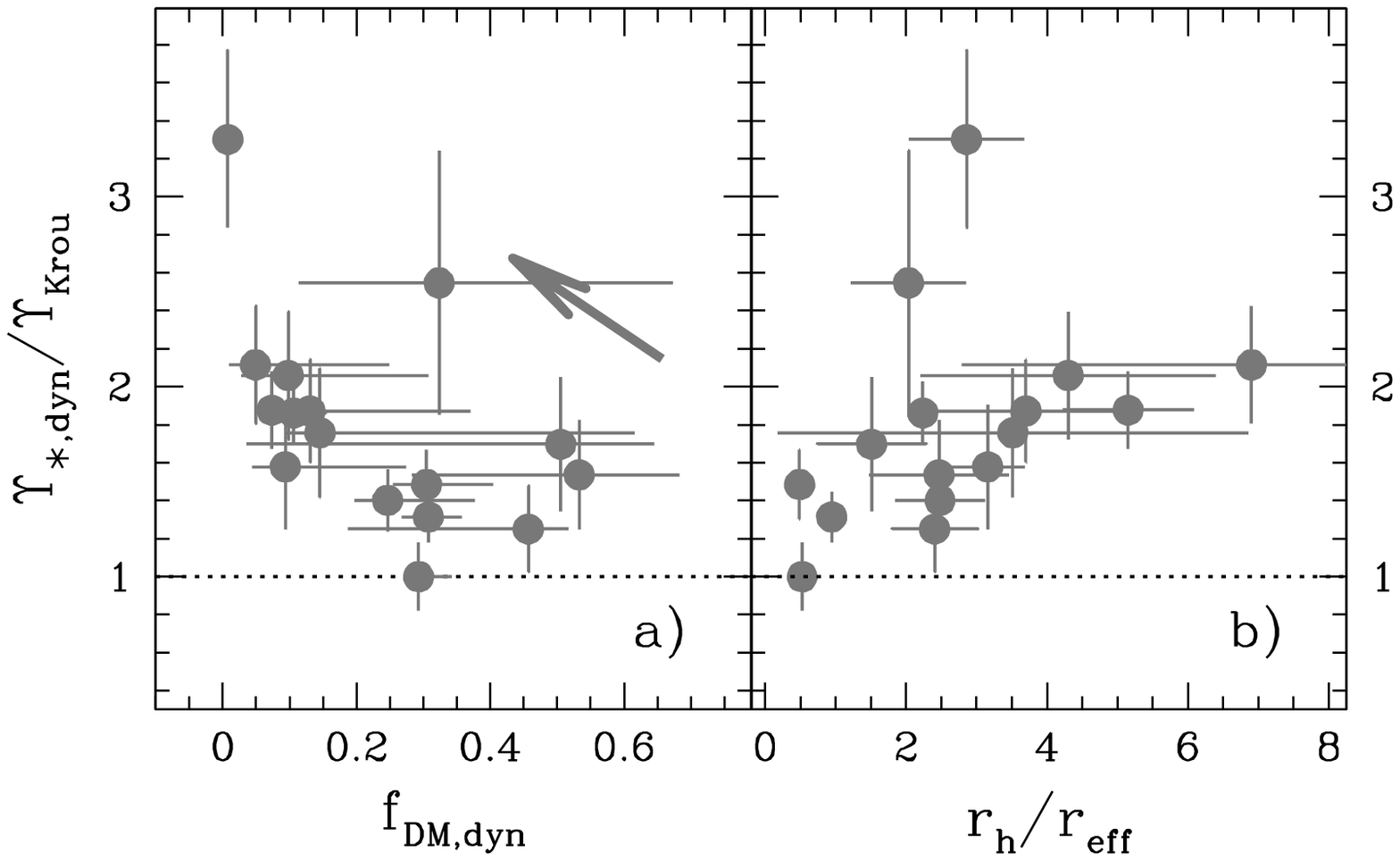}
\caption{$\ratml$ against dark matter fraction $\dmfrac$ (panel a) and
  against halo core radius $\rh$ (scaled by the effective radius
  $\reff$; panel b). Shifting mass from the dark halo component into
  the luminous one at constant {\it total} mass moves galaxies along
  the direction indicated by the arrow in the left-hand panel
  (a). Note that for the right-hand panel the galaxy GMP1990 has been
  omitted, since its dark matter fraction is so low that the
  determination of a halo core-radius becomes meaningless.}
\label{fig:ratml_dm}
\end{minipage}
\end{figure*}

Because the {\it total} (luminous + dark) mass $\mtot$ is well
constrained by the dynamical models (inside the region with
kinematical data), a spurious increase in the luminous mass component
would be accompanied by a corresponding decrease in the nominal dark
matter fraction $\dmfrac$ of the models. More specifically, under the
assumption $\mtot = \mtotgal$, the model parameters ($\mldyn,\dmfrac$)
and the actual galaxy parameters ($\mlgal,\dmfracgal$) would be
related via
\begin{equation}
\label{eq:mldegen}
\mldyn = \mlgal + \left( \dmfracgal - \dmfrac \right) \, \upstot,
\end{equation}
where $\upstot \equiv \mltot$ is the total mass-to-light 
ratio (including dark matter).
A degeneracy in the mass decomposition (at fixed {\it total} mass)
would therefore correlate the offset 
\begin{equation}
\deltaups \equiv \frac{\mldyn - \mlgal}{\mlkroupa}
\end{equation}
in stellar mass-to-light ratios with the offset 
\begin{equation}
\deltadm \equiv \dmfrac - \dmfracgal
\end{equation}
in dark matter fractions as
\begin{equation}
\label{eq:mlshift}
\deltaups = - \frac{\upstot}{\mlkroupa} \deltadm.
\end{equation}

Fig.~\ref{fig:ratml_dm}a shows $\ratml$ against the models' dark matter
fractions $\dmfrac$ (inside $\reff$). There is a slight trend for
$\ratml$ to be particularly large whenever $\dmfrac$ is low.  Note
that an intrinsic dark matter variation from galaxy to galaxy at a
constant IMF scatters galaxies horizontally in
Fig.~\ref{fig:ratml_dm}a, while an IMF variation at constant dark
matter fraction scatters galaxies vertically. The luminous-dark matter
degeneracy discussed above scatters galaxies along the arrow shown in
Fig.~\ref{fig:ratml_dm}a. It marks the direction along which the fitted
($\mldyn,\dmfrac$) are expected to separate from the galaxies'
($\mlgal,\dmfracgal$) according to equation (\ref{eq:mlshift}) and --
for each galaxy -- depends on the ratio $\upstot/\mlkroupa$. For the
arrow in Fig.~\ref{fig:ratml_dm}a we have used the average $\langle
\upstot/\mlkroupa \rangle = 2.39$ over the Coma sample. The
distribution of the majority of Coma galaxies roughly follows the
arrow, in particular below $\dmfrac \la 0.3$.

Fig.~\ref{fig:ratml_dm}a suggests, though does not unambiguously prove,
that the scatter in $\ratml$ could reflect a degeneracy in the
dynamical mass decomposition.  A contamination of some $\mldyn$ with
dark matter would also explain the trend seen in Fig.~\ref{fig:ratml_dm}b,
where $\ratml$ is plotted against the halo core-radius $\rh$ (in units
of $\reff$). The more the inner dark matter goes into $\mldyn$, the
less the model component $\rhodm$ traces the actual inner dark matter
of the galaxies. Instead, it only represents the outer parts of the
dark matter halos. Correspondingly, one would expect relatively larger
halo core-radii (with respect to $\reff$) whenever $\mldyn$ is large
compared to $\mlssp$ -- as seen in Fig.~\ref{fig:ratml_dm}b.

\citet{Nap10} analysed a large set of early-type galaxies with
two-component spherical Jeans models and find dynamical $\mldyn$ close
to the Salpeter IMF when using collisionless halos from cosmological
simulations without baryon contraction. However, their $\mldyn$ get
closer to the Kroupa IMF when baryonic contraction is taken into
account. Baryonic contraction might therefore be one way to make
luminous and dark matter distributions similar enough to explain the
difference between $\mldyn$ and $\mlkroupa$ (see
Sec.~\ref{subsec:degdis}).

In our modelling approach, the halo parameters are allowed to vary
freely, without being connected to results from cosmological
simulations. On the one hand this ensures that baryonic contraction is
implicitly taken into account: the best-fit models for more contracted
halos are simply expected to occur in a different region of parameter
space.  On the other hand, the steepest halo density profiles that we
probed are those from cosmological simulations without baryon
contraction (NFW halos; cf. Sec.~\ref{subsec:dynmod}). If the actual
galaxy halo profiles are steeper, then the best-fit dynamical model
would still be obtained by shifting some fraction of the inner dark
mass into $\mldyn$.

More similar distributions of luminous and dark matter in some
galaxies than in others could also reflect differences in their
evolutionary histories. For example, the cosmological simulations of
\citet{Naa09} indicate a difference in the radial distribution of
in-situ formed stars relative to stars that were accreted during
mergers.  In-situ formed stars have a more centrally concentrated
radial distribution than stars that were accreted in collisionless
mergers. The latter dominate the stellar mass density around
$\reff$. In any case, more detailed investigations of numerical
simulations are required to conclude about a possible degeneracy
between luminous and dark matter in the inner regions of galaxies.

\subsection{The distribution of dark matter in case of a universal Kroupa IMF}
\begin{figure}\centering
\includegraphics[width=84mm,angle=0]{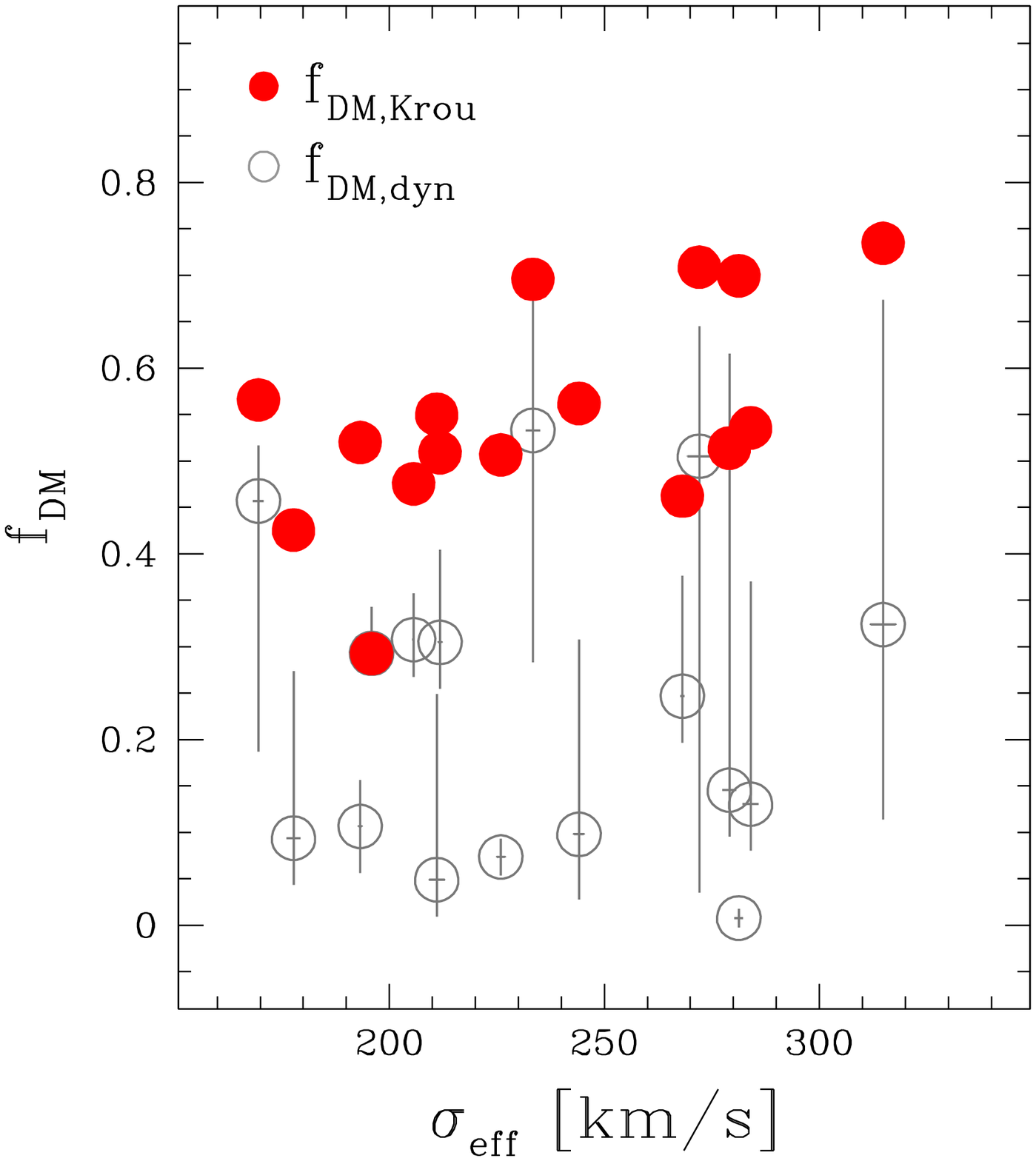}
\caption{Dark matter fractions (inside the effective radius $\reff$)
  against velocity dispersion $\siggal$. Open circles: fraction of dark matter
  $\dmfrac$ that does not follow the light; filled circles: dark matter fraction
  $\dmfrack$ assuming i) a universal Kroupa IMF and ii) that the
  excess mass $(\mldyn-\mlkroupa) \times L$ is a component of dark
  matter that follows the light.}
\label{fig:dmkroupa}
\end{figure}

If we adopt the point of view that the stellar IMF in early-type
galaxies is universal and Kroupa-like then this affects the
distribution of dark matter significantly. The reason is that in this
case our nominal halo component captures only a part of the galaxies'
dark matter, while a large fraction follows the light and is included
in $\mldyn$. In fact, a universal Kroupa IMF implies the dark matter
fractions $\dmfrack$ to read
\begin{equation}
\label{eq:dmfrack}
\dmfrack(r) \equiv \dmfrac(r) + \frac{(\mldyn-\mlkroupa) \times L(r)}{\mtot(r)}.
\end{equation}
These fractions are larger than the nominal $\dmfrac$ derived from
$\rhodm$, have a smaller scatter and slightly increase with
galaxy $\siggal$ (cf. Fig.~\ref{fig:dmkroupa}).

%%%%%%%%%%%%%%%%%%%%%%%%%%%%%%%%%%%%%%%%%%
% f_DM,Kroupa vs M_dyn
%%%%%%%%%%%%%%%%%%%%%%%%%%%%%%%%%%%%%%%%%%

\begin{figure}\centering
\includegraphics[width=84mm,angle=0]{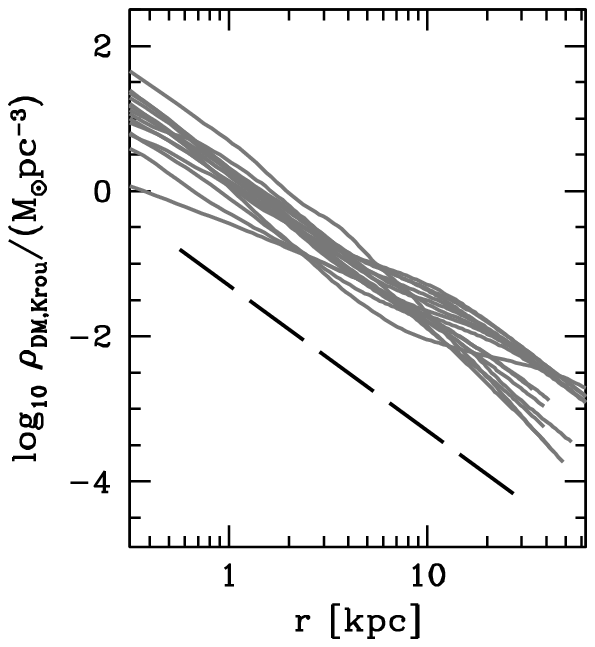}
\caption{Spherically averaged dark matter density $\rhodmk$ against
  radius.  The excess mass $(\mldyn-\mlkroupa) \times L$ with respect
  to a Kroupa IMF stellar population has been added to the dark matter
  halo.  The dashed line indicates a power-law density with
  logarithmic slope of -2.}
\label{fig:rhokroupa}
\end{figure}

Fig.~\ref{fig:rhokroupa} shows the spherically averaged dark matter
density profiles
\begin{equation}
\label{eq:rhodmk}
\rhodmk \equiv \rhodm + (\mldyn-\mlkroupa) \times \nu
\end{equation}
including the extra dark matter required for a Kroupa IMF
(cf. equation \ref{eq:massdecomp}). These density profiles are smooth
and close to a power-law with logarithmic slope slightly shallower
than -2. The slight wiggles in the profiles around $\approx 5 \, \kpc$
indicate the transition from the outer parts, which are dominated by
$\rhodm$, to the inner parts, which are dominated by the second term
on the right-hand side of equation (\ref{eq:rhodmk}).

%%%%%%%%%%%%%%%%%%%%%%%%%%%%%%%%%%%%%%%%%%
% Ups_dyn/Ups_Krou vs sigma
%%%%%%%%%%%%%%%%%%%%%%%%%%%%%%%%%%%%%%%%%%

%%%%%%%%%%%%%%%%%%%%%%%%%%%%%%%%%%%%%%%%%%
% Ups_dyn/Ups_Krou vs sigma
%%%%%%%%%%%%%%%%%%%%%%%%%%%%%%%%%%%%%%%%%%
\begin{figure}\centering
\includegraphics[width=84mm,angle=0]{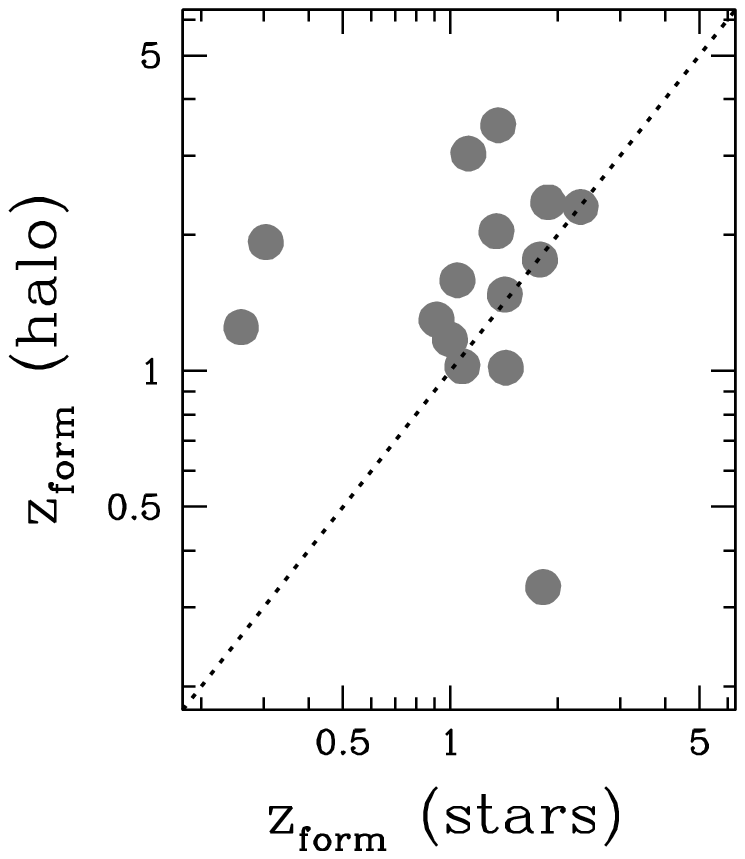}
\caption{Star formation epoch (x-axis) against halo assembly epoch
  (y-axis). The dotted line represents the one-to-one relation.}
\label{fig:zz}
\end{figure}

\subsection{Dark matter density and halo assembly epoch}
\label{subsec:degdis}
In \citet{Tho09} we estimated halo assembly epochs $z_\mathrm{form}$
based on the assumption that the average dark matter density $\langle
\rho_\mathrm{DM} \rangle$ inside $2 \, \reff$ scales with
$(1+z_\mathrm{form})^3$. From the overdensity of dark matter in
early-type relative to spiral galaxies one can then narrow down
elliptical galaxy assembly redshifts with an additional assumption
about the typical formation redshift of spiral galaxies
($z_\mathrm{form} \approx 1$).  Adopting a universal Kroupa IMF
results in average dark matter densities a factor of $\approx 3$
larger than the nominal ones of the dynamical models. The strongest
contribution to this increase comes from the single galaxy GMP1990
which has a negligible $\dmfrac$, but a large $\mldyn/\mlkroupa$.
Without this galaxy, dark matter densities are only larger by about a
factor of $1.6$, which lets the halo assembly redshifts increase from
$z_\mathrm{form} \approx 1 - 3 $ \citep{Tho09} to about
$z_\mathrm{form} \approx 1.5 - 3.5$.

As a second test, we have compared the $\rhodmk$ from
Fig.~\ref{fig:rhokroupa} directly to the galaxy formation models of
\citet{deL07}. These models do not include the dynamical reaction of
dark matter on the baryon infall. Therefore, we first subtracted from
the observed dark matter profile $\rhodmk$ the expected effect of
baryon contraction by assuming the adiabatic approximation (i.e. by
inverting the equations of \citealt{Blu86}). The relationship between
the average dark matter density and halo formation redshift in the
models of \citet{deL07} is well fitted by $\log \langle
\rho_\mathrm{DM} \rangle \approx -2.9 + \log (1+z_\mathrm{form})^3$.
We used this relation to calculate Coma galaxy assembly redshifts from
the decontracted Kroupa dark matter halo densities. Fig.~\ref{fig:zz}
shows that these halo assembly redshifts cover a similar range
($z_\mathrm{form} \approx 1 - 3$) as the ones obtained through the
comparison with spiral galaxy dark matter densities. Note that we here
computed star-formation redshifts from the average stellar ages inside
$\reff$, while in \citet{Tho09} we used the central stellar ages from
\citet{Meh03}.

For many of the Coma galaxies, star formation redshifts and dark halo
assembly redshifts are similar. Two galaxies appear far to the left of
the one-to-one relation. These are GMP0756 and GMP1176, which have
extended and relatively young ($\tau \approx 3 \, \Gyr$) stellar
disks.  The galaxy on the bottom-right, where the stars seem
significantly older than the halo is GMP5568. As discussed in
\citet{Tho09}, dry mergers can move galaxies below the one-to-one
relation in Fig.~\ref{fig:zz}. Besides this possibility the effective
radius of GMP5568 is exceptionally large ($\reff \approx 27 \, \kpc$).
Therefore, the dark matter density refers to a spatial region
significantly larger than in any of our other Coma galaxies. If the
effective radius of this galaxy was overestimated, then the average
dark matter density would be underestimated and so would be the halo assembly
redshift. This could also explain the high dark matter fraction of GMP5568
(cf. Tab.~\ref{tab:dmfrac}), since it generally increases with the physical distance
from the galaxy centre.

\section{The tilt of the fundamental plane}
\label{sec:fp}
The effective radius $\reff$, mean surface brightness $\ieff$ inside
$\reff$ and the central velocity dispersion $\sigma_0$ of a galaxy are
connected via
\begin{equation}
\label{fp1}
\sigma_0^2 = c_M \frac{G \, M}{\reff}
\end{equation}
where $M$ is the total mass and
\begin{equation}
\label{fp2}
\ieff = \frac{L}{2 \pi \, \reff^2}
\end{equation}
where $L$ is the total luminosity. In virial equilibrium the structure
coefficient $c_M$ depends on the orbital structure and the radial
distributions of the mass and the tracer population. For a homologous
family of dynamical objects $c_M$ is a constant and with $\avml \equiv
M/L$ equations (\ref{fp1}) and (\ref{fp2}) lead to
\begin{equation}
\label{eq:FPvir}
\log \frac{\reff}{\kpc} = 2 \, \log \frac{\sigma_0}{\kms}
- \log \frac{\ieff}{\iun} + \gamma
\end{equation}
and 
\begin{equation}
\label{fp3}
\gamma = - \log \frac{\avml}{\mlun} - \log \left( 2 \pi \, \frac{G}{\gun} c_M \right).
\end{equation}
The actually observed fundamental plane (FP; \citealt{Djo87,Dre87}) of
early-type galaxies reads
\begin{equation}
\label{eq:FP}
\log \frac{\reff}{\kpc} = \alpha \, \log \frac{\sigma_0}{\kms}
+ \beta \, \log \frac{\ieff}{\iun} + \gamma
\end{equation}
with $\alpha \ne 2$ and $\beta \ne -1$. It is tilted with respect to
the virial plane of equation (\ref{eq:FPvir}).

%%%%%%%%%%%%%%%%%%%%%%%%%%%%%%%%%%%%%%%%%%
% FP: all galaxies
%%%%%%%%%%%%%%%%%%%%%%%%%%%%%%%%%%%%%%%%%%
\begin{figure*}\centering
\begin{minipage}{166mm}
\includegraphics[width=144mm,angle=0]{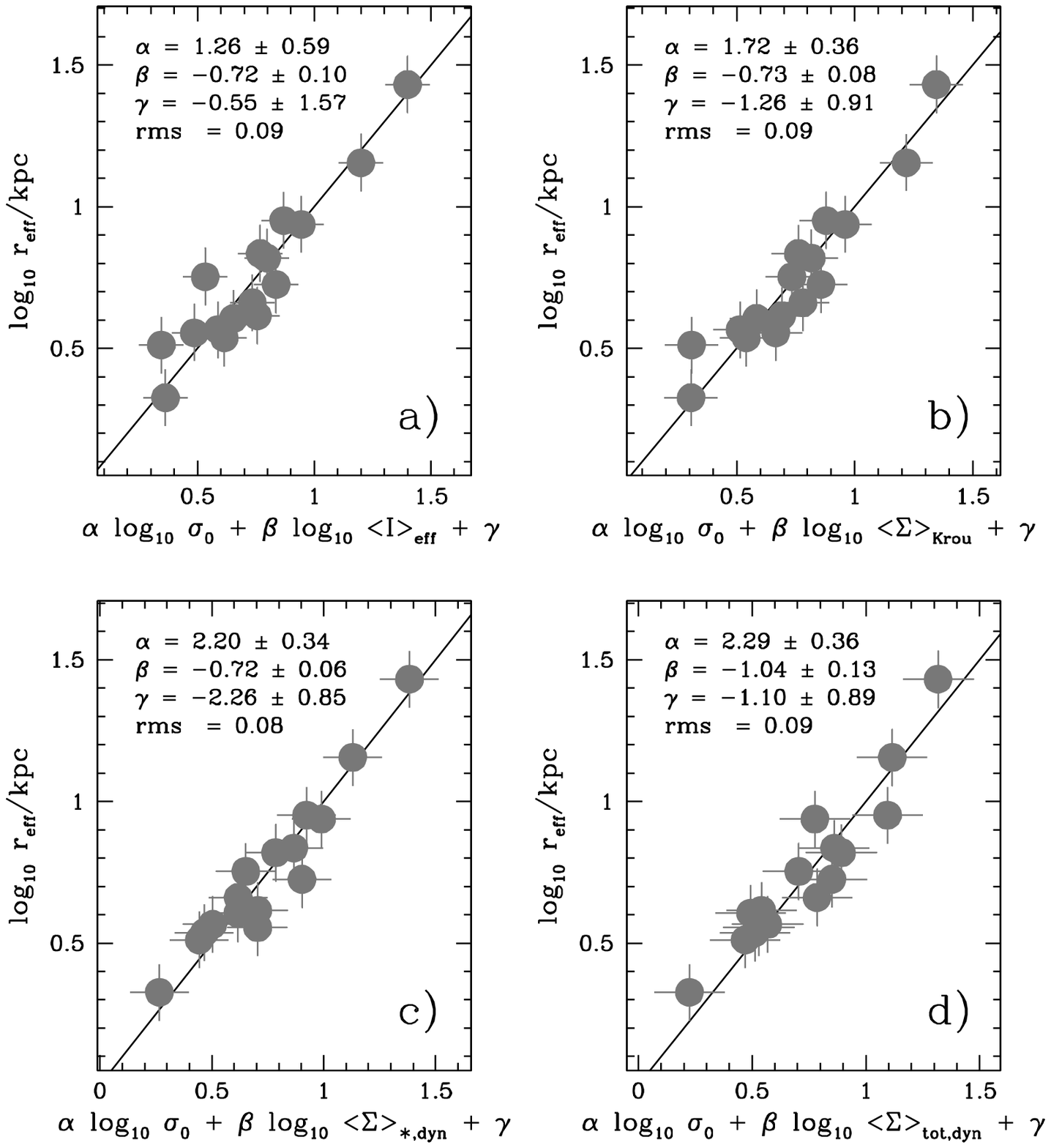}
\caption{Fundamental plane and mass plane of 16 Coma galaxies. In each
  panel the best-fit parameters of an orthogonal fit and the
  rms-scatter in $\log \, \reff$ are quoted.  a) Traditional FP; b)
  fundamental mass plane from stellar populations, i.e. the effective
  surface brightness is $\ieff$ is replaced by the mass-density
  $\sigeff \equiv \mlkroupa \times \ieff$; c) as b) but for the
  dynamically derived stellar-population $\mldyn$: $\sigdyn \equiv
  \mldyn \times \ieff$; d) as c), but instead of the dynamically
  derived {\it stellar} mass-to-light ratio the {\it total} $\mltot$
  (including dark matter) is used: $\sigtot \equiv \mltot \times
  \ieff$ ($\mltot$ is taken at the effective radius). Solid lines
  trace the one-to-one relation.  }
\label{fig:FPall}
\end{minipage}
\end{figure*}

%%%%%%%%%%%%%%%%%%%%%%%%%%%%%%%%%%%%%%%%%%
% FP: old stellar populations
%%%%%%%%%%%%%%%%%%%%%%%%%%%%%%%%%%%%%%%%%%
\begin{figure*}\centering
\begin{minipage}{166mm}
\includegraphics[width=144mm,angle=0]{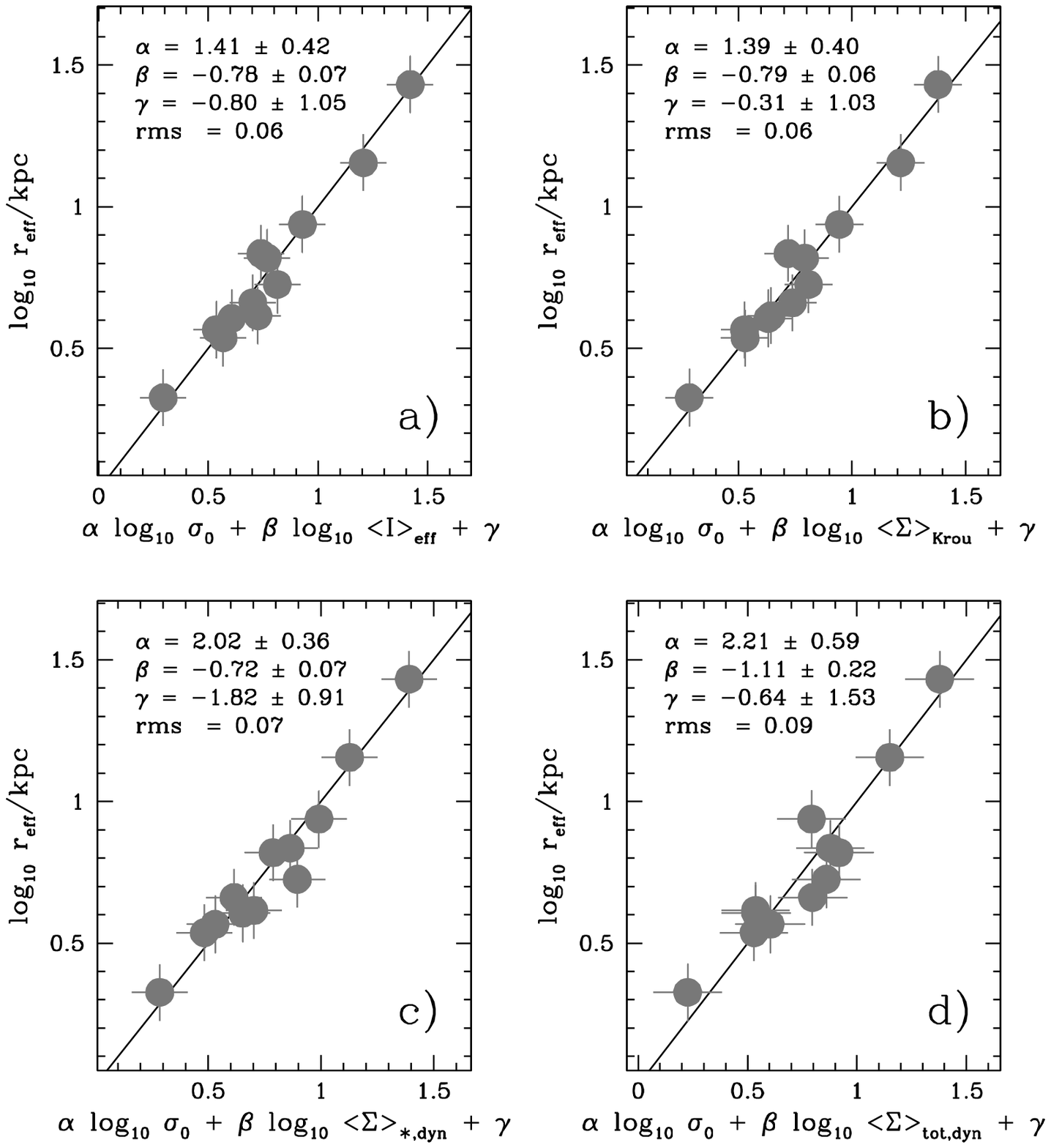}
\caption{As Fig.~\ref{fig:FPall}, but four galaxies with young stellar
  cores are omitted.}
\label{fig:FPyoung}
\end{minipage}
\end{figure*}

Fig.~\ref{fig:FPall}a shows the FP\footnote{Note that in this section
  we use the average velocity dispersion $\sigma_0$ inside the central
  $2\arcsec$, derived in the same way as the effective $\siggal$
  discussed in Sec.~\ref{sec:projmass}.} of Coma galaxies. The
best-fit parameters of an orthogonal fit are $\alpha = 1.26 \pm 0.58$
and $\beta = -0.72 \pm 0.10$ (bootstrap errors).  Within the
statistical uncertainties, the fit is consistent with the $r$-band FP
of SDSS early-type galaxies ($\alpha = 1.49 \pm 0.05$, $\beta = -0.75
\pm 0.01$; \citealt{Ber03,Hyd09}). The larger errors result from the
smaller sample size. Most of the difference with respect to
\citet{Ber03} goes back to four galaxies (GMP0144, GMP0756, GMP1176,
and GMP5975) that are distinct from the rest of the sample in several
respects: (1) they have young central stellar cores ($\tau_0 < 7 \,
\Gyr$; cf. \citealt{Meh03}); (2) at least two of them have an extended
thin stellar disk; (3) they follow different dark halo scaling
relations \citep{Tho09}.  Fig.~\ref{fig:FPyoung} is equivalent to
Fig.~\ref{fig:FPall} except that these four galaxies have been
removed. The corresponding FP matches well with the SDSS results.

\subsection{The fundamental mass plane}
The tilt in the FP can reflect non-homology (i.e. $c_M$ being a
function of galaxy mass) or variations in $\avml$ (or both). A
systematic variation of $\avml$, in turn, could reflect stellar
population effects or changes in the dark matter distribution.  In any
case, the dynamical models allow to incorporate variations of $\avml$
into the FP. For this purpose let $\langle \Sigma
\rangle_\mathrm{eff}$ denote the average surface mass density inside
$\reff$. Then, equation (\ref{eq:FP}) can be written
\begin{equation}
\label{eq:MP}
\log \frac{\reff}{\kpc} = \alpha \, \log \frac{\sigma_0}{\kms}
+ \beta \, \log \frac{\langle \Sigma \rangle_\mathrm{eff}}{\mun}
+ \gamma,
\end{equation}
where 
\begin{equation}
\label{fp4}
\gamma = - \log 
\left( 2 \pi \, \frac{G}{\gun} c_M \right).
\end{equation}
Equation (\ref{eq:MP}) defines the so-called fundamental mass plane
\citep{Bol07}.

Figs.~\ref{fig:FPall}b - \ref{fig:FPall}d show Coma galaxy MPs for
different choices of $\langle \Sigma \rangle_\mathrm{eff}$. Using
either $\langle \Sigma \rangle_\mathrm{Krou} \equiv \mlkroupa \times
\ieff$ (panel b) or $\langle \Sigma \rangle_\mathrm{\ast,dyn} \equiv
\mldyn \times \ieff$ (panel c), the tilt is reduced, but does not
vanish ($\beta \ne -1$).  The change in $\alpha$ from
Fig.~\ref{fig:FPall}a to Fig.~\ref{fig:FPall}b is consistent with the
SDSS analysis of \citet{Hyd09}.  Fig.~\ref{fig:FPall}d shows the case
of $\langle \Sigma \rangle_\mathrm{eff}$ including all the projected
mass (luminous and dark) inside $\reff$, respectively.  The mass plane
of Fig.~\ref{fig:FPall}d is tilt-free within the errors. The match to
the virial plane of equation (\ref{eq:FPvir}) improves further when
omitting the four galaxies harbouring young stellar cores
(cf. Fig.~\ref{fig:FPyoung}).  With $\avml \equiv \mltot$ the MP has
the same scatter as the FP itself (cf. Figs.~\ref{fig:FPall}a and
d). For the subsample of old Coma galaxies the scatter slightly
increases (cf. Fig.~\ref{fig:FPyoung}).

\subsection{The tilt of the fundamental plane}
The tilt in the FP is reduced when the effective surface brightness is
replaced by an effective surface mass derived from
$\mlssp$\footnote{In Sec.~\ref{sec:fp} the surface mass was derived
  from the Kroupa IMF, but the difference between the Kroupa and the
  Salpeter IMF is only a constant scaling factor. The results for the
  Salpeter IMF are therefore similar.}.
The amount of reduction is consistent with the SDSS results of \citet{Hyd09}.
In accordance with \citet{Gra10}, the tilt is further reduced if
$\mlssp$ is replaced by the dynamical $\mldyn$. As discussed above, 
the different scalings of $\mldyn$ and
$\mlssp$ with $\siggal$ could reflect (1) changes in the IMF or (2) changes 
in the distribution of dark matter. The absence of any correlation between $\ratml$ and
stellar population parameters makes (2) more likely than (1). 
Finally, the tilt vanishes completely (for a subsample of Coma 
galaxies with uniformly old stellar populations), if the remaining dark mass inside 
the effective radius $\reff$ is taken into account as well. This behaviour is also found 
in lensing studies \citep{Bol07}. 

In our FP analysis we have not tried to calculate $c_M$ directly from
the dynamical models. However, the density distribution (in particular
the galaxy flattening) as well as the orbital structure vary among the
Coma galaxies \citep{Tho09a}. Nevertheless, Fig.~\ref{fig:FPyoung}d
indicates that for old Coma early-types the tilt of the FP is
dominated by mass-to-light ratio effects rather than any possible
variation in $c_M$.

\subsection{Mass estimators}
\label{subsec:virest}
\begin{figure}\centering
\includegraphics[width=84mm,angle=0]{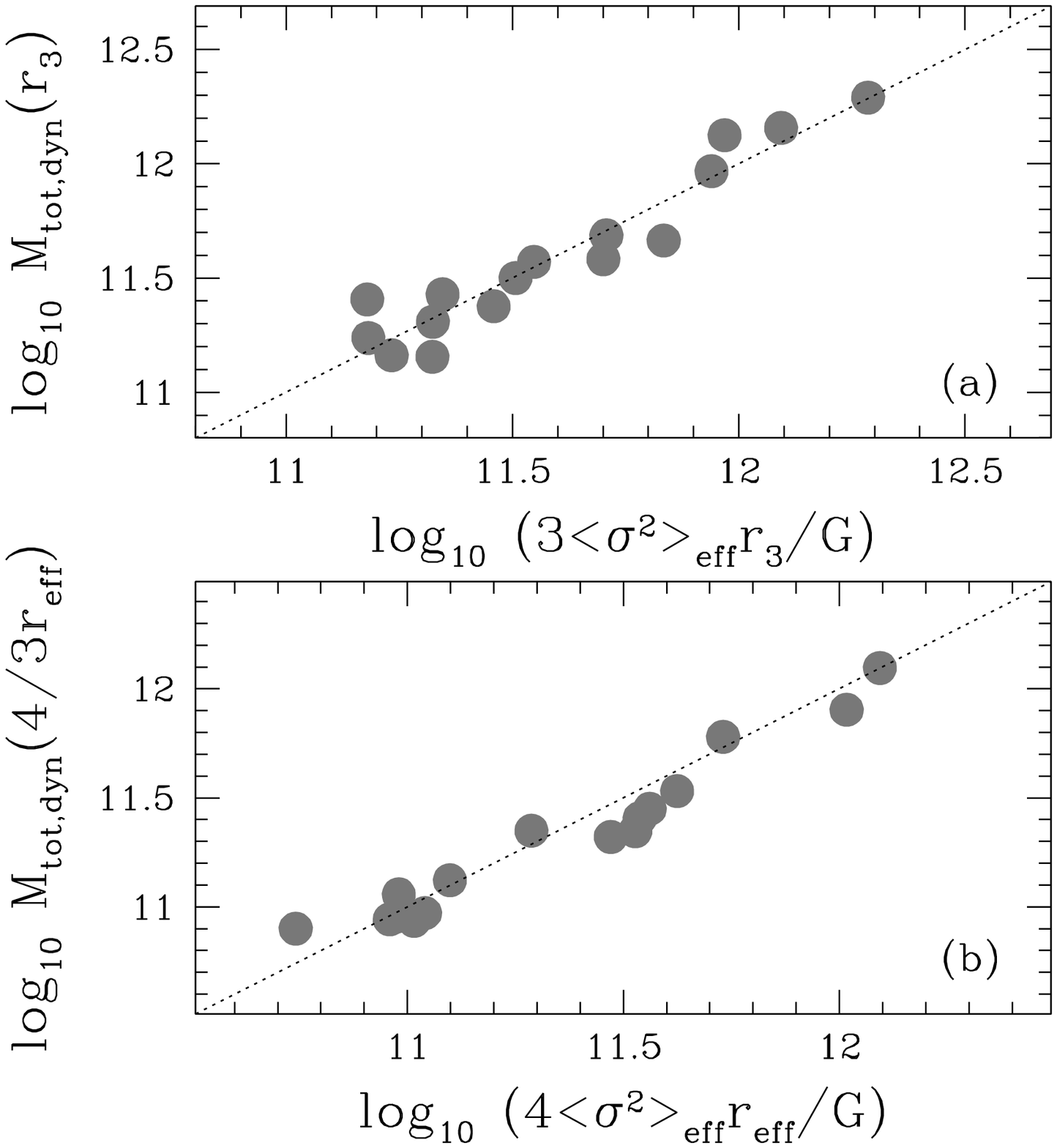}
\caption{Comparison of Coma galaxies with the \citet{Wol10} mass
  estimator (cf. eq.~\ref{eq:Wol1}): (a) total dynamical mass enclosed
  inside the radius $r_3$, where the logarithmic slope of the
  luminosity distribution equals $\xi = -3$; (b) is similar to (a) but
  with the additional approximation $r_3 \approx 4/3 \, \reff$.}
\label{fig:magic_wolf}
\end{figure}

\cite{Wol10} provide an estimation for the mass 
\begin{equation}
\label{eq:Wol1}
M_\mathrm{W10}(r_3) = \frac{3 \langle \sigma^2 \rangle r_3}{G}
\end{equation}
inside the radius $r_3$ where the logarithmic density slope of the
tracer population is $\xi=-3$. Here, $\langle \sigma^2 \rangle$ is the
average of the projected $\sigma^2$ over the whole
galaxy. \citet{Wol10} derived eq.~(\ref{eq:Wol1}) for non-rotating
spherical galaxies. For the Coma galaxies, we averaged the measured
$\sigma^2$ up to $\reff$, ignoring the galaxies' rotation velocities.
Fig.~\ref{fig:magic_wolf}a compares Coma galaxy masses $\mtot(r_3)$
against the predictions of eq.~(\ref{eq:Wol1}). Averaged over all Coma
galaxies we find $\langle \mtot(r_3)/M_\mathrm{W10}(r_3) \rangle =
1.03 \pm 0.27$ (rms scatter). Fig.~\ref{fig:magic_wolf}b is similar to
Fig.~\ref{fig:magic_wolf}a, except for the additional approximation
$r_3 \approx r_{1/2} \approx 4/3 \, \reff$, where $r_{1/2}$ is the
deprojected half-light radius. The agreement with the Coma galaxies is
still very good.

\begin{figure}\centering
\includegraphics[width=84mm,angle=0]{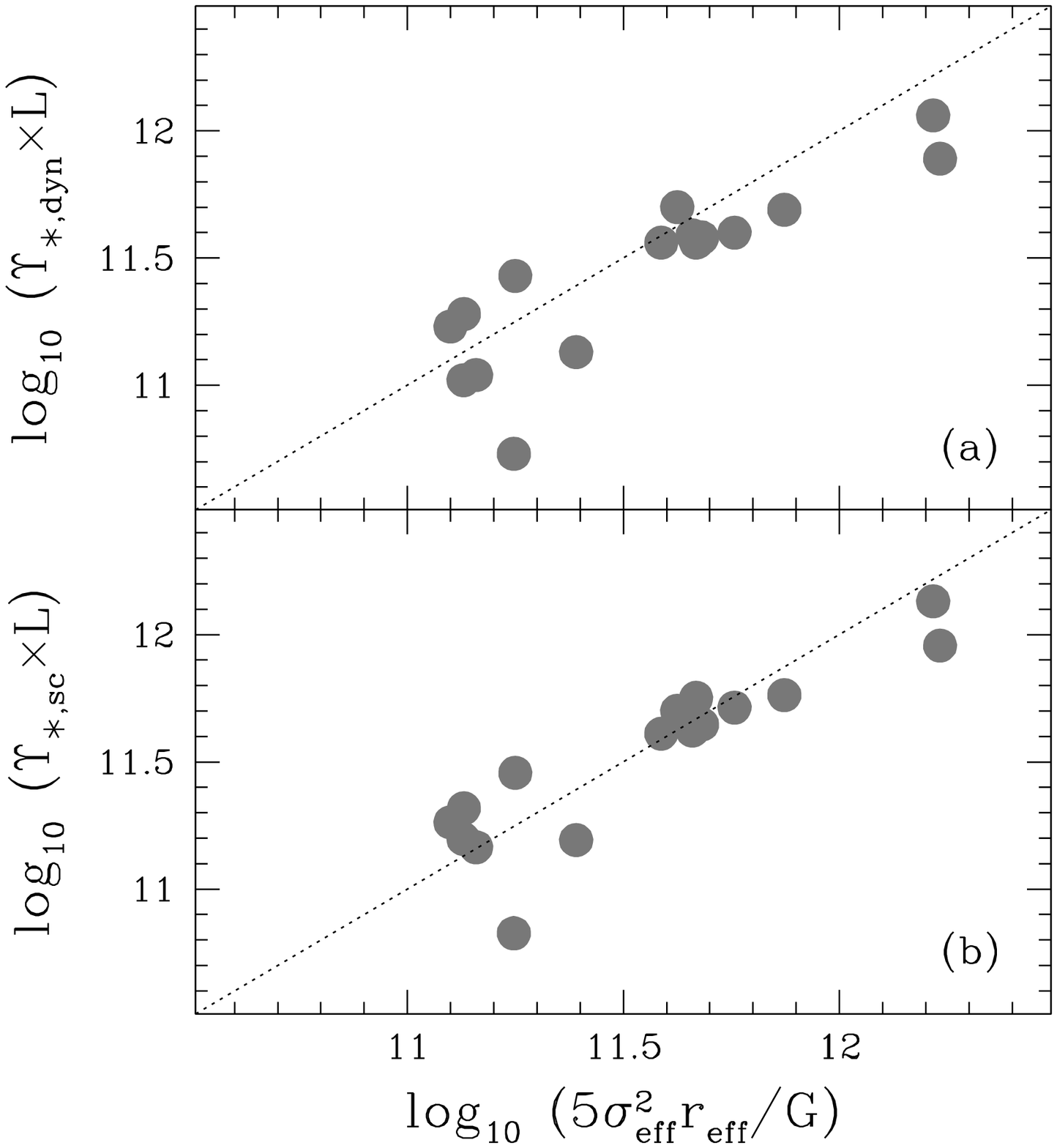}
\caption{Comparison of Coma galaxies with the \citet{Cap06} virial
  mass estimator (cf. eq.~\ref{virialest}): (a) luminous dynamical
  mass $\mldyn \times L$ from fits explicitly allowing for a separate
  component of dark matter; (b) luminous dynamical mass
  $\Upsilon_{\ast,sc} \times L$ from fits assuming that all the mass
  follows the light.}
\label{fig:magic_cap}
\end{figure}

Fig.~\ref{fig:magic_cap} compares luminous dynamical masses with the
estimator
\begin{equation}
\label{virialest}
M_\mathrm{C06} = \frac{5 \sigma^2_\mathrm{eff} \reff}{G}.
\end{equation}
from \citet{Cap06}. When using luminous dynamical masses $\mdyn =
\mldyn \times L$ from model fits that do have a separate dark matter
component, then we find $\langle \mdyn/M_\mathrm{C06} \rangle = 0.86
\pm 0.35$ (cf. Fig.~\ref{fig:magic_cap}a).  The small offset
disappears if we use model fits that do not have an additional dark
matter component (equivalent to the approximation made in
\citealt{Cap06}). The corresponding luminous dynamical masses $\msc =
\mlsc \times L$ (cf. Tab.~\ref{tab:dmfrac}) are slightly larger and
the comparison with eq.~(\ref{virialest}) yields $\langle
\msc/M_\mathrm{C06} \rangle = 1.01 \pm 0.36$
(cf. Fig.~\ref{fig:magic_cap}).  The rms scatter includes both the
measurement errors and anisotropy variations \citep{Tho07B}.  Note,
however, that the assumption that all the mass follows the light is
inconsistent with lensing masses (cf. Fig.~\ref{fig:projmass_tot}).

Thus, the \cite{Wol10} formula gives good estimates for the {\it
  total} dynamical mass inside a radius which is a bit larger than
$\reff$. The virial estimator of \citet{Cap06} captures the entire
dynamical mass that results under the assumption that {\it total mass
  follows light}. Assuming $r_3 \approx r_{1/2}$,
eq.~(\ref{virialest}) implies
\begin{equation}
\label{virialest2}
M_\mathrm{C06}(r_3) \approx M_\mathrm{C06}(r_{1/2}) = \frac{M_\mathrm{C06}}{2} = \frac{2.5 \sigma^2_\mathrm{eff} \reff}{G},
\end{equation}
whereas from eq.~\ref{eq:Wol1} and $r_3 \approx 4/3 \, \reff$ it
follows
\begin{equation}
M_\mathrm{W10}(r_3) \approx \frac{4 \langle \sigma^2 \rangle \reff}{G}.
\end{equation}
Concerning the enclosed mass inside $4/3\reff$, the two mass
estimators would differ by a factor $4/2.5 = 1.6$, unless $\langle
\sigma^2 \rangle \ne \sigma^2_\mathrm{eff}$.  Due to the different
treatment of rotation $\sigma^2_\mathrm{eff} \approx 1.3 \, \langle
\sigma^2 \rangle$ in the Coma sample, such that the actual difference
is only about 20 percent.

%%%%%%%%%%%%%%%%%%%%%%%%%%%%%%%%%%%%%%%%%%
% f_DM vs r_eff
%%%%%%%%%%%%%%%%%%%%%%%%%%%%%%%%%%%%%%%%%%

%%%%%%%%%%%%%%%%%%%%%%%%%%%%%%%%%%%%%%%%%%
% f_DM vs r_eff
%%%%%%%%%%%%%%%%%%%%%%%%%%%%%%%%%%%%%%%%%%

\section{Summary}
\label{sec:sum}
We compared dynamically derived stellar mass-to-light ratios $\mldyn$
with completely independent results from simple stellar population
models. Our dynamical models are based on Schwarzschild's orbit
superposition technique and have two mass components. One follows the
light and its mass-to-light ratio $\mldyn$ is assumed to approximate
the stellar mass distribution. The other mass component explicitly
accounts for dark matter. This way, any potential degeneracy between
the stellar mass and the dark matter halo is minimised.  The Coma
galaxy sample studied here is currently the largest with axisymmetric
Schwarzschild models including dark matter explicitly.

Intrinsic uncertainties in the modelling, in particular related to the
assumption of axial symmetry, are unlikely to bias our results
significantly. The main reason is that in projection, our dynamical
masses match well with completely independent results from strong
gravitational lensing.

Our main findings are:
\begin{enumerate}
\item For galaxies with low velocity dispersions ($\siggal \approx 200
  \, \kms$), the assumption that all the mass follows the light yields
  projected masses larger than in comparable strong-gravitational lens
  systems.
\item In high-velocity dispersion galaxies ($\siggal \approx 300 \,
  \kms$) the assumption that mass follows light is consistent with
  strong lensing results.
\item Two-component dynamical models with an explicit dark halo
  component yield total projected masses that are in good agreement
  with results from strong gravitational lensing for all galaxies.
\item In two-component models, the mass-to-light ratio $\mldyn$ of the
  component that follows the light increases with galaxy velocity
  dispersion $\siggal$.
\item Stellar population $\mlssp$ (for any fixed IMF) are largely
  independent of $\siggal$. As a result, the ratio $\mldyn/\mlssp$ of
  luminous dynamical mass over stellar population mass increases with
  galaxy velocity dispersion.
\item The luminous dynamical $\mldyn$ is always larger than, or at
  least equalises, the stellar-population mass-to-light ratio
  $\mlkroupa$ for a Kroupa IMF.
\item There is no correlation between $\mldyn/\mlssp$ and stellar
  population age, metallicity or [$\alpha$/Fe] ratio.
\item Inside $\reff$, the average fraction of dark matter (that does
  not follow the light) is $\dmfrac = 28 \pm 17 \, \%$ in the Coma
  galaxies.
\item The tilt of the FP reduces if the effective surface brightness
  $\ieff$ is replaced by the stellar population surface-mass density
  $\mlssp \times \ieff$, further reduced if $\ieff$ is replaced by the
  dynamical stellar surface-mass density $\mldyn \times \ieff$ and,
  for a subsample of galaxies with uniformly old stellar populations,
  vanishes completely with the {\it total} dynamical surface-mass
  density $\langle \Sigma_\mathrm{tot,dyn} \rangle_\mathrm{eff}$.
\item Commonly used mass estimators are accurate to the $20 - 30 \,
  \%$ level.
\end{enumerate}

The implications of these findings are as follows:
\begin{enumerate}
\item That luminous dynamical masses increase more rapidly with galaxy
  velocity dispersion than stellar-population masses for a fixed IMF
  could be due to a change in the IMF or due to an increasing amount
  of dark matter following a spatial distribution similar to that of
  the light.
\item If the IMF changes, then massive early-types ($\siggal \approx
  300 \, \kms$) have up to two times more stellar mass per stellar
  light than lower-mass galaxies ($\siggal \approx 200 \, \kms$),
  which are consistent with a Kroupa IMF. However, the lack of any
  correlation between $\mldyn/\mlssp$ and stellar population age,
  metallicity or [$\alpha$/Fe] ratio is consistent with, though does
  not prove, that the IMF is actually universal.
\item If the IMF is universal, then the increase in luminous dynamical
  masses must primarily come from a component of dark matter that
  follows the light very closely and is more important in more massive
  galaxies. The IMF would be consistent with being Kroupa in all
  early-types.
\item Independent of the actual slope of the stellar IMF, luminous
  dynamical masses are {\it on average} more accurately predicted by
  assuming a Salpeter IMF: $\langle \mldyn/\mlsalp \rangle = 1.15$,
  but these masses may not represent exclusively stars. The Kroupa IMF
  yields $\langle \mldyn/\mlkroupa \rangle = 1.8$.
\item Adopting a Kroupa IMF and counting the excess mass
  $(\mldyn-\mlkroupa) \times L$ as dark matter that follows the light
  doubles the average dark matter fractions inside $\reff$ to about
  $\dmfrack \approx 55 \pm 12 \, \%$.  Moreover, it yields a smooth
  trend between the resulting $\dmfrack$ and galaxy velocity
  dispersion and, also, smooth dark matter halo profiles.
\item The FP tilt is not a pure stellar population effect. Further
  inferences about the tilt depend on the interpretation of the
  observed $\ratml$.  As above, that the tilt reduces when considering
  the dynamical mass $\mldyn \times L$ that follows the light could be
  due to (1) variations in the relative distribution of luminous and
  dark matter or (2) IMF variability.
\end{enumerate}

\section*{Acknowledgements}
We thank the referee Glenn van de Ven for useful comments that
helped us to improve the presentation of the results.  JT acknowledges
financial support by the DFG through SFB 375 ``Astro-Teilchenphysik''.
RPS acknowledges support by the DFG Cluster of Excellence ``Origin and
Structure of the Universe''.  EMC is supported by the University of
Padua through grants 60A02-1283/10 and CPDA089220 and by the Ministry
of Education, University, and Research (MIUR) through grant EARA
2004-2006.  SS acknowledges support by the TR33 "The Dark Universe"
and by the Cluster of Excellence ``Origin and Structure of the
Universe''.  The new Coma galaxy HST observations were obtained
through the program HSTGO-10884.0-A which was provided by NASA through
a grant from the Space Telescope Science Institute, which is operated
by the Association of Universities for Research in Astronomy,
Inc. under NASA contract NAS5-26555.  Funding for the SDSS and SDSS-II
has been provided by the Alfred P. Sloan Foundation, the Participating
Institutions, the National Science Foundation, the U.S. Department of
Energy, the National Aeronautics and Space Administration, the
Japanese Monbukagakusho, the Max Planck Society and the Higher
Education Funding Council for England. The SDSS Web Site is
http://www.sdss.org/.  The SDSS is managed by the Astrophysical
Research Consortium for the Participating Institutions. The
Participating Institutions are the American Museum of Natural History,
Astrophysical Institute Potsdam, University of Basel, University of
Cambridge, Case Western Reserve University, University of Chicago,
Drexel University, Fermilab, the Institute for Advanced Study, the
Japan Participation Group, Johns Hopkins University, the Joint
Institute for Nuclear Astrophysics, the Kavli Institute for Particle
Astrophysics and Cosmology, the Korean Scientist Group, the Chinese
Academy of Sciences (LAMOST), Los Alamos National Laboratory, the
Max-Planck-Institute for Astronomy (MPIA), the Max-Planck-Institute
for Astrophysics (MPA), New Mexico State University, Ohio State
University, University of Pittsburgh, University of Portsmouth,
Princeton University, the United States Naval Observatory and the
University of Washington.

\bibliographystyle{mn2e}
%\bibliography{myref}

\appendix

\section{Erratum: Dynamical masses of early-type galaxies: a comparison to
  lensing results and implications for the stellar initial mass function and the
  distribution of dark matter}

The paper 'Dynamical masses of early-type galaxies: a comparison to
lensing results and implications for the stellar initial mass function
and the distribution of dark matter' was published in
Mon. Not. R. Astron. Soc. {\bf 415}, 545-562 (2011). A corrected
version of fig.~5 of that paper is given here along with affected
statements in the text.

\begin{figure*}\centering
\begin{minipage}{166mm}
\includegraphics[width=144mm,angle=0]{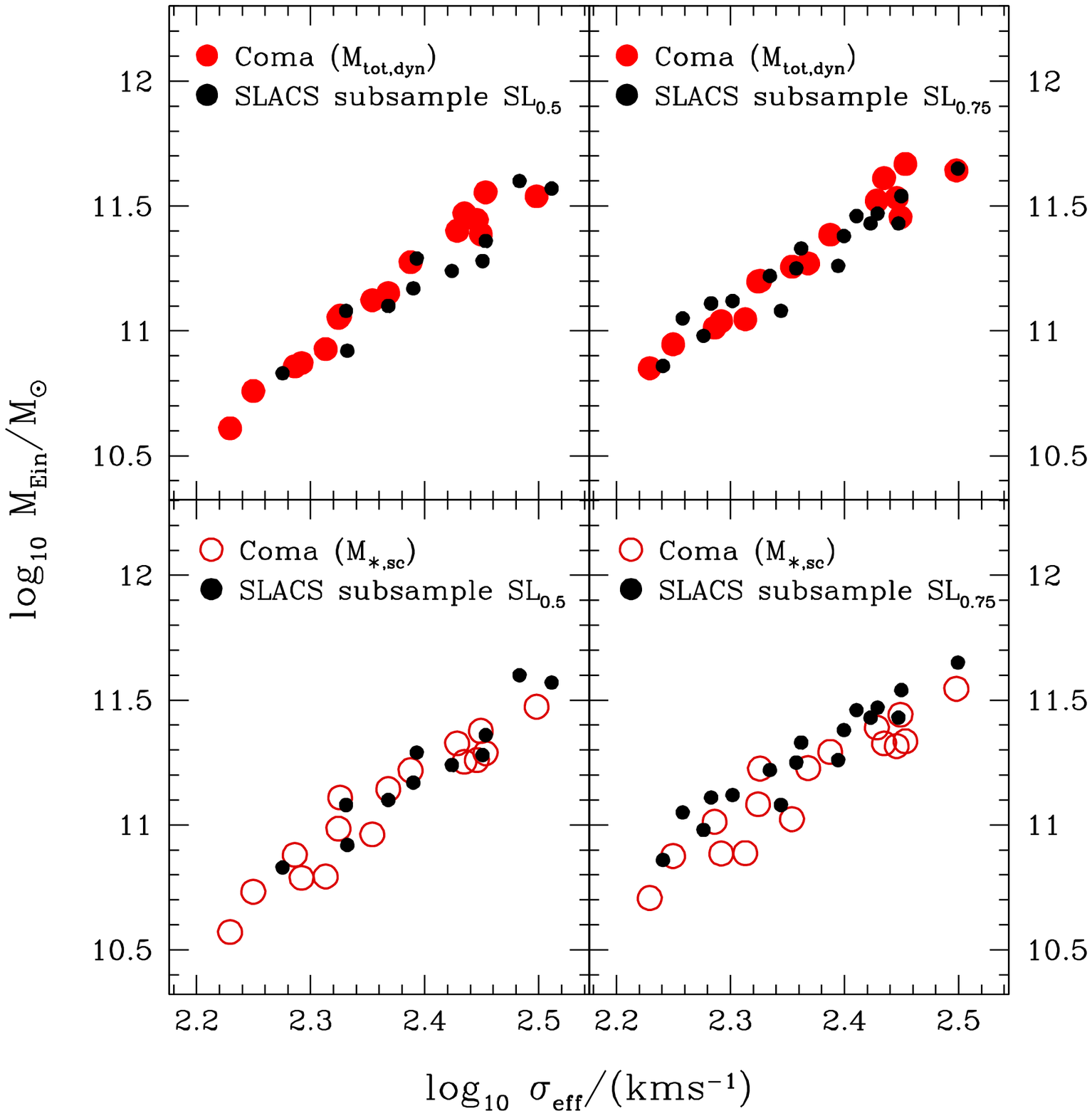}
\caption{Corrected version of fig.~5 from \citet{Tho11}. Top panels
  (as in the original paper): the projected total (luminous+dark) mass
  $\mein$ within a fiducial Einstein radius $\rein$ from two-component
  dynamical models with dark matter haloes. Coma galaxies are indicated
  by the large symbols, while the small circles are SLACS
  gravitational lenses \citep{Aug09}.  Bottom panels: similar
  projected mass, but from dynamical models in which all the mass
  follows the light. In the left-hand panels the comparison is made at
  $\rein \approx 0.5 \, \reff$, while in the right-hand panels at
  $\rein \approx 0.75 \, \reff$ (details in \citealt{Tho11}).  In the
  original version of the paper the Coma galaxies' masses in the {\it
    bottom} panels (open circles) were erroneously raised by 0.19 dex.
}
\label{fig:new}
\end{minipage}
\end{figure*}

In fig.~5 of \citet{Tho11} we compared projected masses of Coma
galaxies obtained (1) from dynamical models with explicit dark matter
halos (top row) and (2) from dynamical models in which all the mass
follows the light (bottom row) with gravitational lenses from the
SLACS survey \citep{Aug09}.  The Coma galaxies' mass estimates without
dark matter ({\it open circles} in the {\it bottom} panels) were
inadvertently raised by 0.19 dex.  Fig.~\ref{fig:new} shows the
correct values. The Coma galaxies' masses of our fiducial models with
dark matter, shown in the upper panels, were correct in the original
paper version and we repeat the upper panels for the sake of
completeness.

Neglecting outer dark matter in dynamical models where all the mass is
assumed to follow the light leads to a mass deficit. Therefore, the
masses obtained from these models fall below those of measured strong
gravitational lenses. The effect becomes noticeable at larger radii
where the luminous mass is less important. This can be seen in the
bottom panels of Fig.~\ref{fig:new}: the left-hand panel is for lenses
which have an Einstein radius $\rein \approx 0.5 \, \reff$ whereas the
right-hand panel is for lenses with $\rein \approx 0.75 \, \reff$. At
$0.5 \, \reff$ projected masses of dynamical models without dark
matter are still consistent with the lenses, while at $0.75 \, \reff$
the mass deficit in the outer parts becomes apparent. Dynamical models
with dark matter (upper panels) are consistent with gravitational lens
masses at all radii.

Consequently, the description of the bottom panels in fig.~5 of
\citet{Tho11}, last paragraph of Sec. 3.2, 'The bottom row of Fig.~5
[...] than in the lower-right one.' is incorrect, while the
conclusion drawn from the comparison -- that dynamical models in which
all the mass follows the light are inconsistent with strong
gravitational lensing results -- remains correct.

In addition, two statements in the Summary (Section~7) are incorrect:
'(i) For galaxies with low velocity dispersions ($\siggal \approx 200
\, \kms$), the assumption that all the mass follows the light yields
projected masses larger than in comparable strong-gravitational lens
systems.' and '(ii) In high-velocity dispersion galaxies ($\siggal
\approx 300 \, \kms$) the assumption that mass follows light is
consistent with strong lensing results.'. Instead, the discrepancy
shows up as a mass deficit and does not depend on the galaxy velocity
dispersion $\siggal$.

We note that in the rest of the paper we only discuss the properties
of our fiducial dynamical models with dark matter and the conclusions
are unaffected by this mistake.

%%%%%%%%%%%%%%%%%%%%%%%%%%%%%%%%%%%%%%%%%%
% Ups_dyn/Ups_Krou vs mass
%%%%%%%%%%%%%%%%%%%%%%%%%%%%%%%%%%%%%%%%%%

\bsp

\label{lastpage}

\end{document}